\documentclass[12pt]{article}

\usepackage{amssymb}
\usepackage{amsmath}
\usepackage{graphicx}
\usepackage[table]{xcolor}
\usepackage[T1]{fontenc}
\usepackage{setspace}
\usepackage[default]{jasa_harvard}    
\usepackage{JASA_manu}
\usepackage[OT1]{fontenc}
\usepackage{rotating}

\usepackage{color}
\newcommand{\add}[1]{\replace{}{#1}}
\newcommand{\remove}[1]{\replace{#1}{}}
\usepackage[normalem]{ulem}
\newcommand{\replace}[2]{\color{red}\textbf{\sout{#1}}\normalcolor\
\color{blue}\textbf{#2}\normalcolor}
\renewcommand{\replace}[2]{#2}

\newcommand{\Acal}{\mathcal{A}}

\newcommand{\ellhat}{\hat{\ell}}

\newcommand{\bfUhat}{\hat{\bfU}}
\newcommand{\bfX}{\boldsymbol{X}}
\newcommand{\bfx}{\boldsymbol{x}}
\newcommand{\bfZ}{\boldsymbol{Z}}
\newcommand{\bfY}{\boldsymbol{Y}}

\newcommand{\bfI}{\boldsymbol{I}}
\newcommand{\bfH}{\mathbf{H}}

\newcommand{\bfWhat}{\hat{\mathbf{W}}}

\newcommand{\bfK}{\mathbf{K}}

\newcommand{\bfSigma}{\mathbf{\Sigma}}

\newcommand{\bftheta}{\boldsymbol{\theta}}
\newcommand{\bfthetahat}{\hat{\boldsymbol{\bftheta}}}
\newcommand{\bfTheta}{\boldsymbol{\Theta}}
\newcommand{\bfomega}{\boldsymbol{\omega}}
\newcommand{\bfmu}{\boldsymbol{\mu}}
\newcommand{\bfomegahat}{\hat{\bfomega}}
\newcommand{\bfOmega}{\boldsymbol{\Omega}}
\newcommand{\omegahat}{ \hat{\omega}}
\newcommand{\sumin}{\sum_{i=1}^n}
\newcommand{\prodin}{\prod_{i=1}^n}
\newcommand{\summM}{\sum_{m=1}^{M}}

\newcommand{\bfU}{\boldsymbol{U}}
\newcommand{\bfV}{\mathbf{V}}
\newcommand{\bfR}{\boldsymbol{R}}

\newcommand{\bfzero}{\mathbf{0}}

\newcommand{\bfy}{\mathbf{y}}

\newcommand{\xsubi}{X^{(i)}}

\newcommand{\ghat}{\hat{g}}

\newcommand{\Uhat}{\hat{U}}

\newcommand{\bfgamma}{\boldsymbol{\gamma}}

\begin{document}

\title{Parsimonious and Efficient Likelihood Composition by Gibbs Sampling}
\author{
Davide Ferrari and Guoqi Qian\\
Department of Mathematics and Statistics, University of Melbourne}
\date{}

\maketitle

\begin{abstract}
The traditional maximum likelihood estimator (MLE) is often of limited use in complex
high-dimensional data due to the intractability of the underlying likelihood function.
Maximum composite likelihood estimation (McLE) avoids full likelihood specification by combining a
number of
partial likelihood objects depending on small data subsets, thus enabling inference for complex
data. A fundamental difficulty in making the McLE approach practicable is the selection from
numerous candidate likelihood objects for constructing the composite likelihood function.
In this paper, we propose a flexible Gibbs sampling scheme for optimal selection of sub-likelihood
components. The sampled composite likelihood functions are shown to converge to the one
maximally informative on the unknown parameters in equilibrium, since sub-likelihood objects are
chosen with probability depending on the variance of the corresponding McLE.  A penalized version
of our method generates sparse likelihoods with a relatively small number of components when the
data complexity is intense. Our algorithms are illustrated through numerical examples on simulated
data as well as real genotype SNP data from a case-control study.
\end{abstract}


\section{Introduction}

While maximum likelihood estimation plays a central role in statistical inference, today
its application is challenged in a number of fields where modern technologies allow
scientists to collect data in unprecedented size and complexity. These fields include
genetics, biology, environmental research,  meteorology and physics, to name a few. Two main
issues arise when attempting to apply traditional maximum likelihood to high-dimensional or complex
data. The first concerns modelling and model selection, since high-dimensional data
typically
imply complex models for which the full likelihood function is difficult, or
impossible, to specify. The second relates to computing, when the full likelihood function
is available but it is just too complex to be evaluated.

The above limitations of traditional maximum likelihood have motivated the development of
 composite likelihood  methods,  which avoid the issues from full
likelihood maximization by combining a set of low-dimensional
likelihood objects. \citeasnoun{Besag74} was an early proponent of
composite likelihood estimation in the
context of data with spatial dependence; \citeasnoun{Lindsay88} developed composite
likelihood inference in its generality and systematically studied its properties. Over the years,
composite likelihood methods have proved useful in a range of complex applications, including models
for geostatistics, spatial extremes and statistical genetics. See
\citeasnoun{Varin&al11} for a comprehensive survey of composite likelihood
theory and applications.

Let $\bfX=(X_1, \dots, X_d)^T$ be a $d \times 1$ random vector with pdf (or pmf)
$f(\bfx| \bftheta_0)$, where $\bftheta_0 \in \bfTheta \subseteq \mathbb{R}^p$, $p \geq 1$, is the
unknown parameter. From independent observations $\bfX^{(1)},\dots, \bfX^{(n)}$, one typically
computes the maximum likelihood estimator (MLE), $\bfthetahat_{mle}$,  by maximizing the
likelihood function $L_{mle}(\bftheta) = \prodin f(\bfX^{(i)}| \bftheta)$.
Now suppose that
complications in the $d$-dimensional pdf (pmf) $f(\bfx| \bftheta)$ make it difficult to specify (or compute)
\replace{$\hat{L}_{mle}(\bftheta)$}{$L_{mle}(\bftheta)$} as the data dimension $d$ grows, but it is relatively easy to
specify (or compute) one-, two-,..., dimensional distributions up to some
order for some functions of $X_1,\dots, X_d$. One can then follow \citeasnoun{Lindsay88}
to estimate $\bftheta$ by the maximum composite likelihood estimator (McLE),
which maximizes the composite likelihood function:
\begin{equation} \label{Lcl}
\replace{\hat{L}}{L}_{cl}(\bftheta) = \prod_{m=1}^{M} \replace{\hat{L}}{L}_m(\bftheta),
\end{equation}
where each $\replace{\hat{L}}{L}_m(\bftheta)$ is a user-specified partial likelihood (or sub-likelihood)
depending on marginal or conditional events on variables.  For example, $\replace{\hat{L}}{L}_m$ can be
 defined using a marginal event
$\{x_m\}$ (marginal composite likelihood), pairs of variables such
as $\{x_1, x_2\}$ (pairwise likelihood), or conditional events like $\{x_1, x_2\}|\{x_1\}$
(conditional composite likelihood). For simplicity, we assume  that $\bftheta$ is common to all
sub-likelihood components, so that any factorization based on a subset of $\{\replace{\hat{L}}{L}_m(\bftheta)$,
$m=1,\dots, M\}$ yields a valid objective function.

Although the composite likelihood approach provides a flexible framework with a sound
theory for making inference about $\bftheta$ in situations involving multivariate data,
there exist at least two challenges hindering the efficiency improvement and feasible
computing of McLE in applications. The first challenge lies with selecting the right
sub-likelihood
components for constructing an informative composite likelihood function. The current
practice of keeping all plausible factors in (\ref{Lcl}) is not well justified in terms
of efficiency relative to MLE, since inclusion of redundant factors can deteriorate
dramatically the variance of the corresponding composite likelihood estimator
(e.g., see \citeasnoun{Cox&Reid04}). A better strategy would be to choose a subset of
likelihood components which are maximally informative on $\bftheta_0$, and drop noisy
or redundant components to the maximum extent. However, little work is found in the
literature in regard to optimal selection of sub-likelihood components.
The second challenge lies with the computational complexity involved in the maximization
of  $\replace{\hat{L}}{L}_{cl}(\bftheta)$, which can go quickly out of reach as $d$ (and $M$) increases.
Particularly, note that computing $\replace{\hat{L}}{L}_{cl}$ involves $M \times N_{ops}(d_{cl})$ operations,
where $N_{ops}(d_{cl})$ is the number of operations for each sub-likelihood component. The
computational burden is exacerbated when $\bfTheta$ is relatively large and the different
sub-likelihood factors $\replace{\hat{L}}{L}_m(\bftheta)$ do not depend on distinct elements of
$\bftheta$. One would like to see this computing complexity be alleviated to a manageable level
by applying parsimonious likelihood composition in the presence of high (or ultra-high) dimensional
data.

Motivated by the aforementioned challenges, in this paper we propose  a new class of
stochastic selection algorithms for optimal likelihood composition. Our method uses Gibbs sampler,
a specific Markov Chain Monte Carlo (MCMC)
sampling scheme to  generate informative -- yet parsimonious -- solutions. The resulting
estimates will converge to the one maximally informative on the target parameter $\bftheta_0$
as the underlying Markov chain reaches equilibrium. This is because sub-likelihood
components generated in the algorithms are drawn according to probabilities determined by
a criterion
minimizing the McLE's variance or its consistent approximation. Theory of unbiased estimating
equations prescribes McLE's asymptotic variance as an optimality criterion, i.e., the
$O_F$-optimality criterion, see \citeasnoun[Ch. 2]{Heyde97}, but such an ideal
objective has
scarce practical utility due to the mathematical complexity and computational cost
involved in evaluating the numerous covariance terms implied by the asymptotic variance expression
\cite{Lindsay&al2011}. To address this issue, we replace the asymptotic variance by
a rather inexpensive jackknife variance estimator computed by a one-step Newton-Raphson iteration.
Such a replacement is shown to work very well based on our numerical study.

Another advantage of our approach is that  proper use of the Gibbs sampler can generate
sparse
composition rules, i.e., composite likelihoods involving a relatively small number of
informative sub-likelihood components. Note that the model space implied by the set of all
available sub-likelihood components can be large, even when $d$ is moderate.
For example, if $d=20$, we have
$2^M=2^{{d}\choose{2}}=2^{190}$ possible composition rules based on pair-wise likelihood
components alone. To cope with such a high-dimensionality, we combine Gibbs sampling with a composite
likelihood stochastic stability mechanism. Analogously to the  stochastic stability selection  proposed by
\citeasnoun{Mein10} in the context of high-dimensional model selection, our approach selects a small
but sufficient number of informative likelihood components through the control of the error rates of
false discoveries.

The paper is organized as follows: In Section \ref{sec2}, we outline the main framework
and
basic concepts related to composite likelihood estimation. We also describe the $O_F$-optimality
criterion and introduce its jackknife approximation. In Section \ref{sec3}, we describe  our core
algorithm for simultaneous likelihood estimation and selection. In Section \ref{sec4}, we
discuss an extension of our algorithm by incorporating the ideas of model complexity penalty and
stochastic stability selection. This leads to a second algorithm for parsimonious likelihoods composition
for high-dimensional data. In Section
\ref{sec5}, we illustrate our methods through numerical examples involving simulated data and real
genetic single nucleotide polymorphism (SNP) data from a breast cancer case-control
study. Section \ref{sec:remarks} concludes the paper with final remarks.

\section{Sparse Composite Likelihood Functions} \label{sec2}

\subsection{Binary Likelihood Composition}

Let $\{ \Acal_1, \dots, \Acal_{M}\}$ be a set of marginal or conditional sample sub-spaces
associated  with \replace{pdfs (or pmfs)}{probability density functions (pdfs)} $f_{m}(\bfx \in \Acal_{m}| \bftheta)$. \add{See
\citeasnoun{Varin&al11} for interpretation and examples of $\Acal_m$.} Given
independent $d$-dimensional
observations $\bfX^{(1)},\dots, \bfX^{(n)} \sim f(\bfx| \bftheta_0)$, $\bftheta_0 \in \bfTheta
\subseteq \mathbb{R}^p$, $p \geq 1$,  we define the  composite log-likelihood function:
\begin{align}  \label{comp_lik}
\ell_{cl}(\bftheta, \bfomega) =
\summM
\omega_{m}
\ell_{m}(\bftheta),
\end{align}
where $\bfomega=(\omega_1, \dots, \omega_{M})^T \in \bfOmega =\{0,1\}^{M}$, and
\replace{$\ellhat_{m}(\bftheta)$}{$\ell_{m}(\bftheta)$} is the partial log-likelihood
\begin{align}
\ell_{m}(\bftheta)  =  \sumin
\log f_{m}(\bfX^{(i)} \in \Acal_m| \bftheta).
\end{align}
Each partial likelihood object (sub-likelihood) $\ell_{m}(\bftheta)$ is allowed to
be selected or not, depending on whether $\omega_m$  is $1$ or $0$. We aim to
approximate the unknown complete log-likelihood function $\ell_{mle}(\bftheta)= \log
\replace{\hat{L}}{L}_{mle}(\bftheta)$ by selecting a few -- yet the most informative --
sub-likelihood objects from the $M$ available objects, where $M$ is allowed to be larger than $n$
and $p$. Given
the composition rule $\bfomega \in \bfOmega$, the maximum composite likelihood estimator (McLE),
denoted by $\bfthetahat(\bfomega)$, is defined by the
solution of the following system of estimating equations:
\begin{align}  \label{comp_score}
\bfzero = \sumin \bfU^{(i)}(\bftheta, \bfomega) :=  \sumin \summM \omega_m
\bfU^{(i)}_{m}(\bftheta),
\end{align}
where
$
\bfU^{(i)}_m(\bftheta) = \nabla_{\bftheta} \log f_m(\bfX^{(i)} \in \Acal_m|\bftheta)
$,
$m=1,\dots,M$, \add{$i=1,\dots,n$}, are unbiased partial scores functions. Under standard regularity conditions on the
sub-likelihoods
\cite{Lindsay88}, $\sqrt{n}(\bfthetahat(\bfomega) - \bftheta_0(\bfomega))
\overset{\mathcal{D}}{\rightarrow}
N_p(\bfzero,
\bfV_0)$ with asymptotic variance given by the  $p \times p$ matrix
\begin{align} \label{asyvar}
\bfV_0(\bfomega) = \bfV(\bftheta_0, \bfomega) = \bfH(\bftheta_0,
\bfomega)^{-1} \bfK(\bftheta_0, \bfomega) \bfH(\bftheta_0,
\bfomega)^{-1},
\end{align}
where
$$
\bfH(\bftheta, \bfomega) =  \summM
\omega_m \bfH_m(\bftheta),
\ \
\bfK(\bftheta, \bfomega) = Var\left[\summM
\omega_{m}
\bfU_{m}(\bftheta) \right],
$$
$\bfH_m(\bftheta) = Var(\bfU_m(\bftheta))$ is the $p\times p$ \replace{Hessian}{sensitivity} matrix for the
$m$th component, and $\bfU_m(\bftheta) =\nabla_{\bftheta} \log f_m(\bfX
\in \Acal_m|\bftheta)$.

The main approach followed here is to minimize, in some sense, the
asymptotic variance, $\bfV_0(\bfomega)$. To this end, theory of unbiased
estimating equations suggests that both matrices $\bfH$ and $\bfK$ should be considered in order to
achieve this goal (e.g., see \citeasnoun[Chapter 2]{Heyde97}). On one hand,
note that $\bfH$ measures the covariance between the composite likelihood score $\bfU(\bftheta,
\bfomega)$\add{$=\sum_{m=1}^M\omega_m
\bfU_m(\bftheta) =\sum_{m=1}^M \omega_m\nabla_{\bftheta} \log f_m(\bfX \in \Acal_m|\bftheta)
$} and the MLE score $\bfU_{mle}(\bftheta) = \nabla_{\bftheta} \log
f(\bfX;\bftheta)$. To see this, differentiate both sides in
\replace{$E[\bfUhat(\bftheta_0,\bfomega)]=\bfzero$}{$E[\bfU(\bftheta_0,\bfomega)]=\bfzero$} and obtain
$
\bfH(\bftheta_0, \bfomega) = \remove{minus sign -}E \left[
\bfU_{mle}(\bftheta_0) \bfU(\bftheta_0, \bfomega)^T \right].
$
This shows that adding sub-likelihood components is desirable, since it increases the
covariance with the full likelihood. On the other hand, including too many correlated
sub-likelihoods components inflates the variance through the covariance terms in
$\bfK(\bftheta_0, \bfomega)$.
\subsection{Fixed-Sample Optimality and its Jackknife Approximation} \label{OF}

The objective of minimizing the asymptotic variance is still
undefined since $\bfV(\bftheta_0, \bfomega)$ in (\ref{asyvar}) is a $p \times p$ positive
semidefinite matrix. Therefore, we consider the following one-dimensional
objective function
\begin{eqnarray} \label{ideal}
\label{g} g_0(\bfomega)=  \log \det \{ \bfV(\bftheta_0, \bfomega) \} = \log \det \{
\bfK(\bftheta_0, \bfomega) \} - 2 \log \det \{ \bfH(\bftheta_0, \bfomega) \}.
\end{eqnarray}
The minimizer, $\bfomega_0$, of the ideal objective (\ref{ideal}) corresponds to the
$O_F$-optimal solution (fixed-sample optimality) (e.g., see \citeasnoun{Heyde97}, Chapter
2). Clearly, such a program still lacks practical relevance, since (\ref{ideal}) depends
on the unknown parameter $\bftheta_0$. Therefore,  $g_0(\bfomega)$ should be replaced by some
sample-based estimate, say $\ghat_0(\bfomega)$.

One option is to use the following consistent estimates of
$\bfH(\bftheta_0, \bfomega)$ and $\bfK(\bftheta_0, \bfomega)$ in
(\ref{g}):
\begin{eqnarray}
\hat{\bfH}(\bfomega) = \dfrac{1}{n-1}\summM \sumin  \bfomega_m
\bfU^{(i)}_m(\bfthetahat)
\bfU^{(i)}_m(\bfthetahat)^T, \ \
\hat{\bfK}(\bfomega) = \dfrac{1}{n} \sumin
\bfU^{(i)}(\bfthetahat, \bfomega)
\bfU^{(i)}(\bfthetahat, \bfomega)^T,
\end{eqnarray}
where the estimator $\bfthetahat=\bfthetahat(\bfomega)$ is the McLE. Although this strategy works
in simple models when $n$ is relatively large and $M$ is small, the estimator
$\hat{\bfK}(\bfomega)$ is knowingly unstable when  $n$ is small compared to
$\dim(\bfTheta)$ \cite{Varin&al11}. Another issue with this approach in  high-dimensional datasets
is that the number of operations required to compute $\hat{\bfK}(\bfomega)$ (or  $\bfK(\bftheta,
\bfomega)$)
increases quadratically in $\summM \omega_m$.

To reduce the computational burden and avoid numerical instabilities, we estimate $g_0$ by the
following one-step \replace{jackknifed}{jackknife} criterion:
\begin{equation}
\ghat(\bfomega) = \log \det \left\{ \sumin \left(
\bfthetahat(\bfomega)^{(-i)} - \overline{\bftheta}(\bfomega) \right) \left(
\bfthetahat(\bfomega)^{(-i)} -  \overline{\bftheta}(\bfomega)\right)^T\right\},
\end{equation}
where the pseudo-value $\bfthetahat(\bfomega)^{(-i)}$ is  a composite
likelihood estimator based on a sample without observation $\bfX^{(i)}$, and
$\overline{\bftheta}(\bfomega)= \sumin \bfthetahat(\bfomega)^{(-i)}/n$. Alternatively, one could
use the delete-$k$ jackknife estimate where $k>1$ observations at the time are deleted to compute
the pseudo-values. The delete-$k$ version is computationally cheaper than
delete-$1$ jackknife and therefore should be preferred when the sample size, $n$, is moderate or
large.
Other approaches -- including bootstrap -- should be
considered depending on the model set up. For example, block re-sampling
techniques such as the block-bootstrap (see \citeasnoun{Hall95} and subsequent papers) are viable
options for spatial data and time series.

When the sub-likelihood scores are in closed form, the pseudo-values can be
efficiently approximated by the following one-step Newton-Raphson iteration:
\begin{equation} \label{pseudovalues}
\bfthetahat(\bfomega)^{(-i)} = \tilde{\bftheta} + \left( \summM
\omega_m \sum_{j \neq i}  \bfU_m^{(j)}(\tilde{\bftheta})
\bfU_m^{(j)}(\tilde{\bftheta})^T \right)^{-1} \left( \summM\omega_m \sum_{j \neq i}
\bfU_m^{(j)}(\tilde{\bftheta}) \right),
\end{equation}
where $\tilde{\bftheta}$ is any root-$n$ consistent estimator of $\bftheta$. Note that
$\tilde{\bftheta}$ does not need to coincide with the McLE, $\bfthetahat(\bfomega)$, so a
computationally cheap initial estimator -- based only on a few small
sub-likelihoods subset  -- may be considered.  Remarkably, the number of
operations required in the Newton-Raphson iteration (\ref{pseudovalues}) grows  linearly in the
number of sub-likelihood components, so that the one-step jackknife objective has
computational complexity comparable to a single
evaluation of the composite likelihood function (\ref{comp_score}). Large sample properties of
jackknife estimator of McLE's asympotic variance can be derived under
regularity conditions on $\bfU_m$ and $\bfH_m$ analogous to those described in
\citeasnoun{Shao92}. Then, for the one-step jackknife using the
root-$n$ consistent starting point $\tilde{\bftheta}$, we
have
$
n \left( \ghat(\bfomega) - g_0(\bfomega) \right) \overset{p}{\rightarrow}
0, \ \ \text{as} \ \ n\rightarrow \infty.
$
uniformly on $\bfOmega$. Moreover, the estimator $\ghat(\bfomega)$ is asymptotically
equivalent to the classic jackknife estimator.

\section{Parameter estimation and likelihood composition} \label{sec3}

\subsection{Likelihood Composition via Gibbs Sampling} \label{sec3.1}

When computing the McLE of $\bftheta$, our objective is to find an optimal binary
vector $\bfomega^\ast$ to estimate
$
\bfomega_0 = {\text{argmin}}_{\bfomega \in \Omega} \
g_0(\bfomega),
$
where $g_0(\cdot)$ is the ideal objective function defined in (\ref{ideal}).
Typically, the population quantity $g_0$ cannot be directly assessed, so we replace $g_0$
with the sample-based jacknife estimate $\ghat$ described in Section \ref{sec2} and aim at
finding
$$
\bfomega^\ast = \underset{\bfomega \in \Omega}{\text{argmin}} \
\ghat(\bfomega).
$$
This task, however, is computationally
infeasible through enumerating the  space $\bfOmega$ if $d$ is even moderately large.
For example, for composite likelihoods  defined based on all pairs of variables  $(X_s, X_t)$,
$1\leq s<t\leq d$ (pairwise likelihood),  $\bfOmega$ contains $2^{{d}\choose{2}}=2^{190}$ elements
\add{when $d=20$.}

To overcome this enumeration complexity,  we carry out a random search method based on Gibbs
sampling. We regard the weight vector $\bfomega$ as a random vector following the joint probability
mass function (pmf)
\begin{equation} \label{Gibbs_prob}
\pi_\tau(\bfomega)  = \dfrac{1}{Z(\tau)}\exp \left\{-\tau \ghat(\bfomega)
  \right\},  \ \ \bfomega \in \bfOmega,
\end{equation}
where $Z(\tau)= \sum_{\bfomega \in \bfOmega} \exp \{-\tau  \ghat(\bfomega) \}$ is
the normalizing constant. The above distribution
depends on the tuning parameter $\tau>0$,
which controls the extent to which we emphasize larger probabilities (and reduce smaller
probabilities) on $\bfOmega$. Then $\bfomega^*$ is also the mode of $\pi_\tau(\bfomega)$, meaning
that
$\bfomega^*$ will have the highest probability to appear, and will be more likely to appear earlier
rather than later, if a random sequence of $\bfomega$
is to be generated from $\pi_\tau(\bfomega)$.

Therefore, estimating $\bfomega^*$ can be readily done based on the random sequence generated from
$\pi_\tau(\bfomega)$. But generating a random sample from $\pi_\tau(\bfomega)$ directly is difficult
because $\pi_\tau(\bfomega)$  contains an intractable normalizing constant $Z(\tau)$. Instead, we
will generate a Markov chain using the product of all univariate conditional pmf's with respect to
$\pi_\tau(\bfomega)$
as the transitional kernel. The stationary distribution of such a Markov chain can be
proved to be $\pi_\tau(\bfomega)$ \cite[Chapter 10]{Casella04}. Hence, the part of this Markov chain
after reaching equilibrium can
be regarded as a random sample from $\pi_\tau(\bfomega)$ for most purposes. The MCMC
method just described is in fact the so-called Gibbs sampling method. The key factor
for the Gibbs sampling to work is all the univariate conditional probability distributions
of the target distribution can be relatively easily simulated.

Let us write
$\bfomega=(\omega_1,\cdots,\omega_M)$; $\bfomega_{m_1:m_2}=(\omega_{m_1},
\omega_{m_1+1},\cdots,\omega_{m_2})$, if $m_1\leq m_2$, and
$\bfomega_{m_1:m_2}=\emptyset$ otherwise; and $\bfomega_{-m}=(\omega_1,\cdots,
\omega_{m-1},\omega_{m+1},\cdots,\omega_M)$. Then it is easy to see that the conditional
pmf of $\omega_m$ given $\bfomega_{-m}$, $m=1,\cdots,M$, is
\begin{equation}
\pi_\tau(\omega_m|\bfomega_{-m})=\frac{\exp\{-\tau \ghat(\bfomega)\}}{
\exp\{-\tau \ghat(\bfomega_m^{[0]})\}+
\exp\{-\tau \ghat(\bfomega_m^{[1]})\}},\quad
\omega_m=0,1,
\label{eq7}
\end{equation}
where
$\omega_m^{[j]}=(\omega_{1:(m-1)},j,\omega_{(m+1):M})$, $j=0,1$. Note that
$\pi_\tau(\omega_m|\bfomega_{-m})$ is simply a Bernoulli pmf and so it is easily
generated. For
each
binary vector $\bfomega$,   $\ghat(\bfomega)$ is computed using the
one-step jackknife estimator described in Section \ref{sec2}. Therefore, the
probability mass function $\pi_\tau(\bfomega)$ is well defined for any $\tau>0$. On the other hand,
we have shown that the
Gibbs sampling can be used to generate a Markov chain from $\pi_\tau(\bfomega)$, from which we can
find a consistent estimator $\hat{\bfomega}^*$ for the mode $\bfomega^*$ \add{if $\bfomega^*$
is unique}. Consequently,
the universally maximum composite log-likelihood estimator of $\bftheta$ can be
approximated by the McLE, $\bfthetahat({\bfomegahat}^*)$.

\add{Note that the mode $\bfomega^*$ is not necessarily unique. In this case one can still consider
consistent estimation for $\bfomega^*$  but in the following meaning. We know
$\ghat(\bfomega^*)$ is the minimum and always unique according to its definition. Thus
$\ghat(\bfomega^*)$ can be consistently and uniquely estimated based on the Markov chain
of $\ghat(\bfomega)$ induced from the Markov chain of $\bfomega$ generated from
$\pi_\tau(\bfomega)$ when the length of the Markov chain goes to infinity. Therefore, any estimator
$\bfomegahat^*$ such that $\ghat(\bfomegahat^*)$ becomes consistent with
$\ghat(\bfomega^*)$ can be regarded as a consistent estimator of $\bfomega^*$.
Consequently, the McLE $\bfthetahat({\bfomegahat}^*)$ for each such $\bfomegahat^*$ is
still the universally McLE but not necessarily unique. From a practitioner's view point
there is no need to identify all consistent estimators of $\bfomega^*$ and all universally McLE
of $\bftheta$. Finding out one such $\bfomegahat^*$ or a tight superset of it would be a
sufficient advance in parsimonious and efficient likelihood composition.}

\subsection{Algorithm 1: MCMC Composite Likelihood Selection (MCMC-CLS)}
\label{sec3.2}

The above discussion motivates the steps of our core Gibbs sampling  algorithm for simultaneous
composite likelihood estimation and selection. Let $\tau$ be given and fixed.

\begin{enumerate}
\item[0$ $.]
For $t = 0$, choose an initial binary vector of weights \add{$\bfomega^{(0)}$} and compute the one-step jackknife
estimator $\ghat(\bfomega^{(0)})$. \add{E.g. randomly set 5 elements of $\bfomega^{(0)}$ to 1 and the rest to 0.}
\item[1$ $.]
For each $t=1,\cdots, T$ for a given $T$, obtain $\bfomega^{(t)}$
by repeating $1.1 $ to $1.4$ for each $m=1,\cdots,M$ sequentially.
\begin{enumerate}
\item[1.1$ $.]
Compute, if not available yet, $\ghat(\omega_{1:(m-1)}^{(t)}, j,\omega_{(m+1):M}^{(t-1)})$,
$j=0,1$.
\item[1.2$ $.]
Compute the conditional pmf of $\omega_m$ given
$(\omega_{1:(m-1)}^{(t)},\omega_{(m+1):M}^{(t-1)})$:
\begin{equation}
\pi_\tau(\omega_m=j|\omega_{1:(m-1)}^{(t)},\omega_{(m+1):M}^{(t-1)})\propto
\exp\{-\tau \ghat(\omega_{1:(m-1)}^{(t)}, j,\omega_{(m+1):M}^{(t-1)})\}
\end{equation}
where $j=0,1$. Note this is a Bernoulli pmf.
\item[1.3$ $.]
Generate a random number from the Bernoulli pmf obtained in $1.2 $, and denote the result
as $\omega_m^{(t)}$.
\item[1.4$ $.]
Set $\bfomega^{(t)}\leftarrow
(\omega_{1:(m-1)}^{(t)},\omega_m^{(t)},\omega_{(m+1):M}^{(t-1)})^T$. Also compute and
record
$\ghat(\bfomega^{(t)})$.
\end{enumerate}
\item[2$ $.] Compute
$ \bfomegahat^*=\arg\min_{1\leq t\leq T}
\ghat( \bfomega^{(t)}),$
and regard it as the estimate of
$\bfomega^*$. Alternatively, column-combine $\bfomega^{(1)},\cdots, \bfomega^{(T)}$ generated in
Step $1$ into an $M\times T$ matrix
$\bfWhat$; then compute row averages of $\bfWhat$,
say $\overline{\omega}_1,\cdots,\overline{\omega}_M$, and set
$\tilde{\bfomega}^*=(\tilde{\omega}_1^*,\cdots, \tilde{\omega}_M^*)$
where
$\tilde{\omega}_m^*=1$, if $\overline{\omega}_m\geq \xi$, and $\tilde{\omega}_m^*=0$
if $\overline{\omega}_m< \xi$, where $\xi$ is some constant larger than $0.5$.
\item[3$ $.] Finally, compute  $\bfthetahat(\hat{\bfomega}^*)$ (or
$\bfthetahat(\tilde{{\bfomega}}^*)$) and $\ghat(\hat{\bfomega}^*)$ (or
$\ghat(\tilde{{\bfomega}}^*)$).
\end{enumerate}
Firstly, note that Gibbs sampling has been used in various contexts in the literature
of model selection. \citeasnoun{George&McCulloch97} used a similar strategy to generate
the distribution of the  variable indicators in Bayesian linear regression.
\add{\citeasnoun{Qian1999} used Gibbs sampler for selecting robust linear regression models.}
\citeasnoun{Qian&Field02} \remove{have} used the
Gibbs sampler for  selecting  generalized linear regression models. \citeasnoun{Brooks&al03} and
\citeasnoun{Qian&Zhao07} \replace{use}{used} Gibbs sampling for selection in the context of time series
models.
However, to our knowledge this is the first work proposing a general-purpose Gibbs sampler for
construction of  composite likelihoods.

Secondly, the sequence $\bfomega^{(1)}, \dots, \bfomega^{(T)}$ is a Markov chain, which
requires \add{an initial vector $\bfomega^{(0)}$ and} a
burn-in period to be in equilibrium. \add{Values of $\bfomega^{(0)}$ do not affect the eventual
attainment of equilibrium so
can be arbitrarily chosen. From a computational point of view most components of $\bfomega^{(0)}$
should be set to 0 to reduce computing load. For example, we can randomly set all but 5 of the components to 0.}
To assess whether the chain has reached equilibrium, we suggest
the control-chart method discussed in \citeasnoun{Qian&Zhao07}.
For the random variable $\ghat({\bfomega})$, we have the
following probability inequality for any given $b>1$:
$$
Pr\left(\ghat(\bfomega) - \min \ghat(\bfomega) \geq b \sqrt{Var[\ghat(\bfomega) ] +
(E[\ghat(\bfomega)] - \min \ghat(\bfomega))^2 }    \right)  \leq \dfrac{1}{b^2}
$$
This inequality can be used to find an upper control limit for $\ghat({\bfomega})$. For example, by
setting \replace{$b=10$}{$b=\sqrt{10}$}, an at least $90\%$ upper control limit for $\ghat({\bfomega})$
can be estimated as
$
\ghat^\ast + \sqrt{10 s^2 + 10 (\overline{g} -\ghat^\ast)^2 },
$
where $\ghat^\ast$, $\overline{g}$  and $s^2$ are the minimum, sample mean and sample variance
based on the first $N$ observations, $\ghat(\bfomega^{(1)}), \dots, \ghat(\bfomega^{(N)})$,
$N < T$, where  typically we set $N=\lfloor T/2 \rfloor$. We then count the number of observations
passing the
upper control limit in the remaining sample  $\ghat(\bfomega^{(N+1)}), \dots,
\ghat(\bfomega^{(T)})$. If more than 10\% of the points are above the control limit, then
at a significance level not more than 90\% there is statistical evidence against
equilibrium. Upper control limits of different levels for $\ghat({\bfomega})$ can be similarly
calculated and interpreted by choosing different values of $b$.

Thirdly, $\bfomegahat^*$ computed in Step~2 is simply a sample mode of $\pi_\tau(\bfomega)$
based on its definition in (\ref{Gibbs_prob}). Hence, by the Ergodic Theorem for stationary Markov
chain, $\bfomegahat^*$ is a strongly consistent estimator of $\bfomega$ under minimal
regularity conditions. With similar arguments $\overline{\omega}_m$ is a strongly consistent
estimator of the success probability involved in the marginal distribution of $\omega_m$
induced from $\pi_\tau(\bfomega)$. Hence, it is not difficult to see that the resultant
estimator $\tilde{\bfomega}^*$ should satisfy $\tilde{\omega}_m^*\geq \omega_m^*$
without requiring $T$ to be very large. Propositions~1 and 2 in \citeasnoun{Qian&Zhao07}
provide an exposition of this property. Therefore, the estimator $\tilde{\bfomega}^*$
captures all informative sub-likelihood components with high probability.

Finally, the tuning constant $\tau$ adjusts the mixing behavior of the chain, which has important
consequences on the exploration/exploitation  trade-off on the search space $\bfOmega$. If $\tau$
is too small, the algorithm produces solutions approaching the global optimal value
$\bfomega^\ast$ slowly; if $\tau$ is large, then the algorithm finds local optima and may
not reach $\bfomega^\ast$. The former behavior corresponds to a rapidly mixing
chain, while the latter occurs when the chain is mixing too slowly. In the composite likelihood
selection setting, the main hurdle is the computational cost, so $\tau$ should be set according to
the available computing capacity, after running some graphical or numerical diagnostics
(e.g., see \citeasnoun{Casella04}). We choose to use $\tau=d$ in our empirical study, which does
not seem to create adverse effects.

\section{An extension for high-dimensional data} \label{sec4}

\subsection{Sparsity-enforcing penalization} \label{sec:penalty}

Without additional modifications, Algorithm 1  ignores the likelihood complexity, since
solutions with many sub-likelihoods have in principle the same chance to occur  as those with fewer
components. To discourage selection of overly complex likelihoods, we augment the Gibbs distribution
(\ref{Gibbs_prob}) as follows:
\begin{equation} \label{augmentedgibbs}
\pi_{\tau, \lambda}(\bfomega) = Z(\tau, \lambda)^{-1} \exp\{ -\tau \ghat_\lambda(\bfomega)\},
\end{equation}
where
\begin{equation} \label{g_penalized}
\ghat_\lambda(\bfomega) = \ghat(\bfomega) + {pen}(\bfomega|\lambda), \quad \tau>0, \;
\lambda>0,
\end{equation}
$ \ghat(\bfomega)$ is the jackknifed variance objective defined in Section \ref{sec2},
$Z(\tau, \lambda)$ is the normalization constant, and $pen(\bfomega)$ is
a complexity penalty enforcing sparse solutions when  $\dim(\bfOmega)$ is large.
Maximization of $\pi_{\tau,
\lambda}(\bfomega)$ is interpreted as a maximum a posteriori estimation for $\bfomega$, where the
probability distribution proportional to $\exp\{ -pen(\bfomega|\lambda)\}$
is regarded as a prior pmf over $\bfOmega$. In this paper, we use
the penalty term of form $\text{pen}(\bfomega|\lambda) = \lambda \summM \omega_m$, since it
corresponds to
well-established model-selection criteria. For example, choices $\lambda=1$, $\lambda = 2^{-1}\log
n$ and $\lambda=\log \log n$ correspond to the AIC, BIC and HQC criteria, respectively (e.g., see
\citeasnoun{Claeskens08}). Other penalties could be considered as well depending on the model
structure and
available prior information; however, these are not shown to be crucial based on our
empirical study
so will not be explored in this paper.

\subsection{Composite Likelihood Stability Selection} \label{sec:stab}

To find the optimal solution $\bfomega^\ast$, one could compute a sequence of optimal values
$\bfomegahat^\ast_{\lambda_1}, \dots, \bfomegahat^\ast_{\lambda_B}$ and then take
$\min_{1 \leq b \leq B} \ghat(\bfomegahat^\ast_{\lambda_b})$. There are, however, various issues in
this approach: first, the
globally optimal value $\bfomega^\ast$  might not be a member
of the set $\{ \bfomegahat^\ast_{\lambda_b} \}_{b=1}^B$, since the mode of
$\pi_{\tau,\lambda}(\bfomega)$ is not necessarily the composite likelihood solution
which minimizes $\ghat(\bfomega)$. Second, even if $\bfomega^\ast$ is in
such a set, determining  $\lambda$ is typically
challenging.  To address the above issues, we employ the idea of stability
selection, introduced by \citeasnoun{Mein10} in the context of variable selection for linear
models. Given an arbitrary value for $\lambda$, stochastic stability exploits the variability of
random
samples generated from $\pi_{\tau, \lambda}(\bfomega)$ by the Gibbs procedure, say
$\bfomega^{(1)}_\lambda, \dots,
\bfomega^{(T)}_\lambda$ and choses all the partial likelihoods that occur in a large fraction
of generated samples. For a given $0<\xi<1$, we define the set of stable likelihoods
by the vector
$\bfomegahat^{\text{stable}}$, with elements
\begin{equation}
\omegahat_m^{\text{stable}} =
\left\{
\begin{array}{ll}
1, &\text{if }  \ \ \dfrac{1}{T}\sum_{t=1}^T
\omega_{\lambda,m}^{(t)}\geq \xi, \\
0, & \text{otherwise},
\end{array}
\right.
\end{equation}
so we regard as stable those sub-likelihoods selected more frequently and disregard sub-likelihood
items with low selection probabilities. Following \citeasnoun{Mein10}, we choose the tuning constant
$\xi$ using the following bound on the expected number of false selections, $V$:
\begin{equation} \label{EV}
E(V) \leq \dfrac{1}{(2\xi-1)} \dfrac{\eta_{\lambda}}{M},
\end{equation}
where $\eta_{\lambda}$ is average number of
selected sub-likelihood components. In multiple testing, the quantity $\alpha
= E(V)/M$ is sometimes called the per-comparison error rate (PCER). By increasing $\xi$, only few
likelihood components are selected, so that we reduce the expected number of falsely selected
variables.  We
choose the threshold $\xi$ by fixing the PCER at some desired value (e.g.,
$\alpha = 0.10$), and then choose $\xi$ corresponding to the desired error rate. The unknown
quantity $\eta_\lambda$ in our setting can be estimated by the average number of sub-likelihood
components over $T$ Gibbs samples.

Finally, note that tuning $\xi$ according to (\ref{EV}) makes redundant the determination of the
optimal $\lambda$ value as long as $pen(\bfomega|\lambda)$ in (\ref{g_penalized}) is not dominant
over $\ghat({\bfomega})$. This is further supported by our empirical study where we found the effect
of $\lambda$ on $\hat{\bfomega}^{\text{stable}}$ is negligible.

\subsection{Algorithm 2: MCMC Composite Likelihood with Stability Selection
(MCMC-CLS2)} \label{alg2}

The preceding discussions lead to Algorithm 2 which is essentially the same as  Algorithm 1 with  two
exceptions: (i) we replace $\ghat$ in Algorithm 1 by the augmented objective function $\ghat_{\lambda}$
defined in (\ref{g_penalized}); (ii) Steps 2 in Algorithm 1 is replaced by the
following stability selection step.
\begin{itemize}
\item[$2^\prime$.]
Column-combine  $\bfomega^{(1)},\cdots, \bfomega^{(T)}$ generated in Step $1$ into an $M\times T$
matrix $\bfWhat$. Compute the row averages of $\bfWhat$, denoted as
$(\omegahat_1,\cdots,\omegahat_M)$, i.e. $\omegahat_m=T^{-1}\sum_{t=1}^T\omega_m^{(t)}$,
$m=1,\cdots, M$. Then set $\bfomegahat^{\text{stable}}=(\omegahat_1^*,\cdots,\omegahat_M^*)$ where
$\omegahat_m^*=1$ if $\omegahat_m\geq \hat{\xi}_\lambda$ and $\omegahat_m^*=0$ if
$\omegahat_m<\hat{\xi}_\lambda$,
$m=1,\cdots,M$, where
$
\hat{\xi}_\lambda= \dfrac{1}{2}\left( \dfrac{\hat{\eta}_\lambda}{\alpha M^2} +
1\right),
$
and $\alpha$ is a nominal level for per-comparison error rate (e.g., 0.05 or 0.1).
\end{itemize}
The estimated threshold $\hat{\xi}_\lambda$ is obtained from (\ref{EV}) by plugging-in
$\hat{\eta}_{\lambda} =  \sum_{t=1}^T \sum_{m=1}^M \omega^{(t)}_m / T$, the
sample average of the \replace{number}{numbers} of selected sub-likelihood components in $T$ Gibbs samples.

\section{Numerical Examples} \label{sec5}

\subsection{Normal Variables with Common Location} \label{sec5.1}

Let $\bfX \sim N_d(\mu \bf1,\bfSigma)$, where the
parameter of interest is the common location \remove{parameter,} $\mu$. We study the scenario where many
components bring redundant
information on $\mu$ by considering covariance matrix $\bfSigma$ with elements
$\{\bfSigma\}_{mm} = 1$, for all $1\leq m\leq d$, and off-diagonal elements
$\{\bfSigma\}_{lm}=\rho>0$ if $l,m \leq d^\ast$, for some
$d^\ast < d$, while $\{\bfSigma\}_{lm}=0$ elsewhere.

We consider one-wise score composite likelihood estimator solving \replace{$0 = \sum_{m=1}^d
\omega_{m}\Uhat_{m}(\mu)$}{$0 = \sum_{m=1}^d
\omega_{m}U_{m}(\mu)$}, where \replace{$\Uhat_m(\mu) = \sumin
(X^{(i)}_m - \mu)$}{$U_m(\mu) = \sumin (X^{(i)}_m - \mu)$}.
\replace{Then}{It is easy to find} the composite likelihood estimator is
$
\widehat{\mu}_{cl}(\bfomega) = {\sum_{m=1}^d \omega_m \overline{X}_m
}/{\sum_{m=1}^d \omega_m }
$
where $\overline{X}_m = \sumin X_m^{(i)}/n$. For this simple model, the \replace{jackknifed}{jackknife} criterion
$\ghat(\cdot)$ can be easily computed in closed form. The pseudo-values are
$\widehat{\mu}^{(-i)}_{cl}(\bfomega) = \sum_{m=1}^d
\omega_m
\overline{X}^{(-i)}_m/\sum_{m=1}^d \omega_m$, and the average of pseudo-values is
$\widehat{\mu}_{cl}$. \replace{This implies}{It can be shown that}
$$
\ghat(\bfomega) = \log \sumin \left( \sum_{m=1}^d \omega_m (X_m^{(i)} - \overline{X}_m) \right)^2
- 2 \log \left( \sum_{m=1}^d \omega_m \right),
$$
 up to a constant not depending on $\bfomega$. \replace{The}{It can also be shown that}
$O_F$-criterion has the following expression
\begin{align}\label{varex}
g_0(\bfomega) = \log Var\left(\widehat{\mu}_{cl}(\bfomega) \right)  \propto
\log \left(\sum_{m=1}^d \omega_m + 2 \rho \sum_{ l < m \leq
d^\ast
} \omega_{l}\omega_{m}  \right) - 2\log\left( \sum_{m=1}^d \omega_m\right),
\end{align}
depending on the unknown parameter $\rho$. This \replace{shows}{suggests} that
including too many correlated (redundant) components (with $\rho
\neq 0$) damages McLE's variance as $d$ increases.
Particularly, setting $\omega_j =1$, for all $j=1,\dots, d$,
implies $Var(\widehat{\mu}_{cl}(\bfomega) ) = O(1)$, while
choosing only uncorrelated sub-likelihoods ($\omega_j=0$ if $2\leq j \leq d^\ast$, and
$\omega_j=1$
elsewhere), gives
$
Var(\widehat{\mu}_{cl}(\bfomega)) = O(d^{-1})
$.

In Table 1, we show Monte Carlo simulation results from $B=250$ runs of Algorithm 1 (MCMC-CLS1) using
the two approaches described in Section \ref{sec3.2}: one
consists of choosing the best weights vector (CLS1 min); the other uses thresholding, i.e. selects elements when
the weights are selected with a sufficiently large frequency (CLS1 thres.). In the same table we also report
results on the Algorithm 2 (MCMC-CLS2) based on the stochastic stability selection approach.
Our algorithms are compared with the estimator including all one-wise sub-likelihood components (No selection)
and the optimal maximum likelihood estimator (MLE). \replace{We consider MLE with optimal weights based on the true
matrix $\Sigma^{-1}$, and  the MLE based on covariance estimator
$\hat{\Sigma} = (n-1)^{-1} \sum_{i=1}^n (X_i- \overline{X})(X- \overline{X})^T$.}{We compute the
MLE of $\mu$ in two ways: either based on using the known $\Sigma$ value or based on using the
sample covariance $\hat{\Sigma} = (n-1)^{-1} \sum_{i=1}^n (X_i- \overline{X})(X_i- \overline{X})^T$.}
Note that the latter is \replace{note}{not} available when $d>n$. \replace{Instead}{Note}  our
algorithms are \remove{mainly} designed to obtain simultaneous estimation and dimension reduction in the
presence of limited information (i.e. $d>n$ or $d \gg n$) \replace{and}{where the}
optimal weights are difficult to estimate from the data. Stochastic selection with $\xi=0.7$ and $T=10d$ was
carried out for samples of $n=5, 25$ and $100$
observations from a model with $d^\ast =0.8 \times d$ correlated components, with $d=10, 30$.
To compare the methods we computed Monte Carlo estimates \replace{for}{of} the
variance ($Var$) and squared bias ($Bias^2$) \add{of $\widehat{\mu}_{cl}(\bfomega)$.}
Table 2 \replace{,}{further} reports the average number of selected
likelihood components (no. comp).

For all considered data  dimensions, our selection methods outperform the all one-wise likelihood (AOW
\add{or No selection})
estimator and show relatively small losses in terms of mean squared error ($Var + Bias^2$)
compared to \replace{MLE with optimal smooth weights.}{the MLEs.} The gains in terms of
variance reduction are particularly evident for larger data dimensions (e.g., see $d=30$). At the
same time, our method tends to select mostly uncorrelated sub-likelihoods, while
discarding the redundant components which do not contribute useful information on $\mu$.

\begin{sidewaystable}
\begin{tabular}{llllllllllllllll}
  \hline
     &     &        & \multicolumn{2}{c}{No selection} & \multicolumn{2}{c}{CLS1 (min)} & \multicolumn{2}{c}{CLS1 (thresh.)} & \multicolumn{2}{c}{CLS2}&
\multicolumn{2}{c}{MLE (known $\Sigma$)} &
\multicolumn{2}{c}{MLE (unknown $\Sigma$)}\\
 $n$ & $d$ & $\rho$ & $Var$ & $Bias^2$&   $Var$ & $Bias^2$&   $Var$ & $Bias^2$&   $Var$ & $Bias^2$&  $Var$ & $Bias^2$&   $Var$ & $Bias^2$&\\
  \hline
 5 & 10 &  0.50 &  80.26  & 7.19   & 79.18 & 0.04 & 81.36 & 0.00 & 87.55 & 0.01 & 52.52 & 0.04 & NA & NA \\
   &    &  0.90 &  115.06 & 10.31  & 96.97 & 0.05 & 114.82 & 0.08 & 114.90 & 0.02 & 62.28 & 0.04 & NA & NA \\
   & 30 &  0.50 &  80.08  & 7.18   & 60.61 & 1.60 & 54.06 & 1.18 & 52.16 & 0.74 & 26.22 & 0.03 & NA & NA \\
   &    &  0.90 &  103.50 & 9.28   & 61.27 & 0.01 & 44.14 & 0.01 & 44.68 & 0.00 & 24.93 & 0.02 & NA & NA \\
   \\
25 & 10 &  0.50 &  17.14 & 8.28  & 15.27 & 0.18 & 17.45 & 0.06 & 17.73 & 0.06 & 12.01 & 0.03 & 20.99 & 0.03 \\
   &    &  0.90 &  19.89 & 1.78  & 14.75 & 0.07 & 18.65 & 0.03 & 19.08 & 0.02 & 13.02 & 0.02 & 21.64 & 0.04 \\
   & 30 &  0.50 &  11.87 & 1.06  & 7.38 & 0.12 & 6.14 & 0.00 & 6.07 & 0.00 & 4.86 & 0.00 & NA &  NA\\
   &    &  0.90 &  24.51 & 2.20  & 11.59 & 0.02 & 6.69 & 0.00 & 6.75 & 0.00 & 5.73 & 0.01 & NA & NA \\
   \\
100& 10 &  0.50 &  4.14 & 0.37  & 3.09 & 0.02 & 3.55 & 0.03 & 4.24 & 0.09 & 2.56 & 0.02 & 2.79 & 0.02 \\
   &    &  0.90 &  6.18 & 0.55  & 3.25 & 0.00 & 4.58 & 0.00 & 4.58 & 0.00 & 3.12 & 0.00 & 3.37 & 0.00 \\
   & 30 &  0.50 &  3.40 & 0.30  & 1.79 & 0.00 & 1.60 & 0.00 & 1.60 & 0.00 & 1.20 & 0.00 & 1.69 & 0.00 \\
   &    &  0.90 &  5.54 & 0.50  & 2.60 & 0.01 & 1.68 & 0.01 & 1.68 & 0.01 & 1.41 & 0.01 & 2.03 & 0.01 \\
   \hline
\end{tabular}
\caption{Bias and variance of estimators for the location model $\bfX \sim N_d(\mu
\bf1,\bfSigma(\rho))$ by different methods. MCMC-CLS1 algorithm with and without thresholding (CLS1 (min) and  MCMC-CLS1 (thresh.), respectively);  MCMC-CLS2
algorithm with stability selection (CLS2) with $\alpha=0.1$ and $\lambda = 1$; Maximum likelihood estimator
\remove{with optimal weights} based on $\Sigma^{-1}$ where
$\Sigma$ is known; Maximum likelihood estimator \remove{with optimal weights} based on $\hat{\Sigma}^{-1}$, where $\hat{\Sigma}$ can be estimated if $d<n$, otherwise
the value is missing and denoted by NA. For each method we show the finite sample variance ($Var$) and squared bias ($Bias^2$) for $n=5,25,100$, $d=10,30$ and
$\rho=0.5,0.9$. Estimates based on $B=250$ Monte Carlo runs (simulation settings: $\tau =  d$, $T=10d$ , $\xi = 0.7$). Monte Carlo standard errors are
smaller than $0.001$.}
\end{sidewaystable}

\begin{table}[h]
\centering
\begin{tabular}{lllcccc}
  \hline
$n$ & $d$ & $\rho$ & \multicolumn{1}{c}{No selection}& \multicolumn{1}{c}{CLS1 (min)}& \multicolumn{1}{c}{CLS1 (thres.)} & \multicolumn{1}{c}{CLS2} \\
  \hline
 5 & 10 & 0.50 & 10 & 3.77 & 3.92 & 3.72 \\
    &  & 0.90 & 10 & 3.28 & 2.58 & 2.36 \\
    & 30 & 0.50 & 30 & 13.94 & 11.01 & 10.84 \\
   &  & 0.90 & 30 & 12.04 & 6.53 & 6.43 \\
   \\

   25 & 10 & 0.50 & 10 & 4.30 & 3.58 & 3.26 \\
   &  & 0.90 & 10 & 3.27 & 2.04 & 2.00 \\
  & 30 & 0.50 & 30 & 12.81 & 7.70 & 7.48 \\
  &  & 0.90 & 30 & 11.96 & 5.98 & 5.98 \\
  \\

 100 & 10 & 0.50 & 10 & 4.28 & 2.56 & 2.31 \\
  &  & 0.90 & 10 & 3.17 & 2.00 & 2.00 \\
  & 30 & 0.50 & 30 & 12.22 & 6.08 & 6.05 \\
  &  & 0.90 & 30 & 11.94 & 6.00 & 6.00 \\
   \hline
\end{tabular}
\caption{Number of selected sub-likelihoods by different methods. MCMC-CLS1 algorithm with and without thresholding (CLS1 (min) and  MCMC-CLS1 (thresh.),
respectively);  MCMC-CLS2
algorithm with stability selection (CLS2) with $\alpha=0.1$ and $\lambda = 1$. For each method we show results for $n=5,25,100$, $d=10,30$ and  $\rho=0.5,0.9$.
Estimates based on $B=250$ Monte Carlo runs (simulation settings: $\tau =  d$, $T=10d$ , $\xi = 0.7$). Monte Carlo standard errors are smaller than $0.01$.}
\end{table}

Next, we illustrate the selection procedure based on MCMC-CLS2 (Algorithm 2). We draw a  random
sample of $n=50$ observations from the model $N_d(\mu \bf1,\bfSigma(\rho))$ described above with
$d=250$,
$d^\ast = 0.8 d$, and $\rho=0.9$. This corresponds to 200 strongly redundant variables and 50
independent variables. We applied Algorithm~2  with $\alpha=0.1$ and $\lambda = 1$ (corresponding to
the AIC penalty).  In Figure 1 (b), we show the relative frequencies of the components,
$\overline{\omega}_m = \sum_{t=1}^T \omega^{(t)}_m$, $m=1,\dots,250$, in $T=1000$ MCMC
iterations. As expected, the uncorrelated sub-likelihood components (components 201--250) are
sampled much more frequently than the redundant ones (components 1--200). Figure 1
(c) shows the objective function $\ghat_\lambda(\bfomegahat^{\text{stable}})$ (up to a constant)
evaluated at the best solution computed from past MCMC samples ($\xi =
0.7$). Figure 1 (d) shows the Hamming distance, $dist(\bfomega,
\bfomega^\prime) = \sum_j I(\omega_j \neq \omega_j^\prime)$, between the current selected rule,
$\bfomegahat^\ast$ and true optimal value, $\bfomega^\ast =
(1,\underbrace{0,\dots,0}_{199},\underbrace{1,\dots,1}_{50})$. Overall, our algorithm
quickly approaches the optimal solution and the final selection has 94.0\% asymptotic
relative efficiency (ARE) compared to MLE. When no stochastic selection is applied and
all 250 sub-likelihood components are included, the relative efficiency is only 13\%. As far as computing time is concerned, our non-optimized R implementations of the MCMC-CLS1 and MCMC-CLS2 algorithms for the above example takes, respectively 4.25 and 1.87 seconds per MCMC iteration, respectively. The computing time was measured on a laptop computer with Intel \circledR  Core$^{\text{TM}}$ i7-2620M CPU \@ 2.70GHz.

\begin{figure}[h]\label{fig1}
\centering
\begin{tabular}{cc}
\hspace{1cm}(a) & \hspace{1cm}(b) \\
 \includegraphics[scale=0.4]{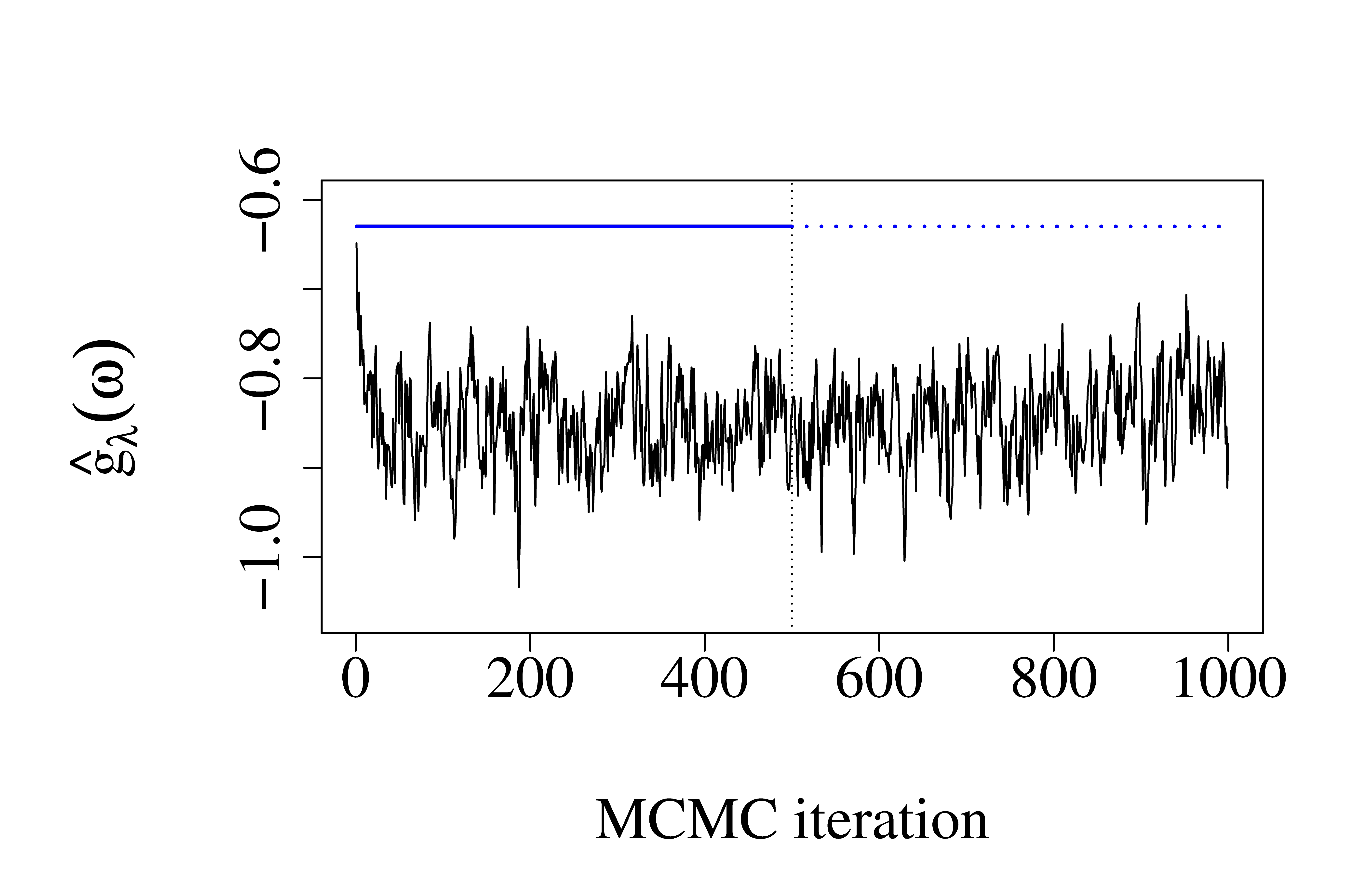}
&\includegraphics[scale=0.4]{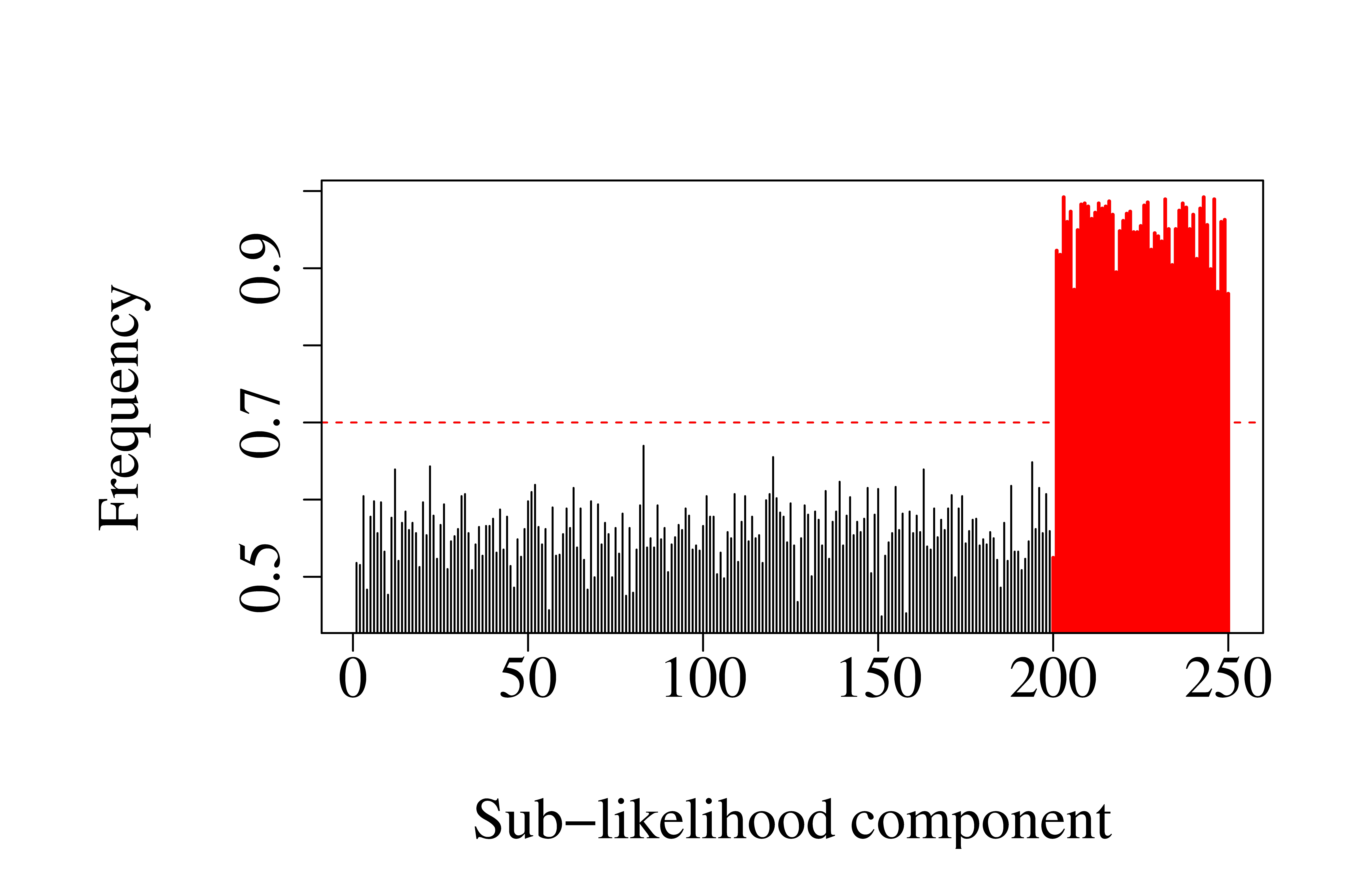}\\
\hspace{1cm}(c) &\hspace{1cm} (d) \\
 \includegraphics[scale=0.4]{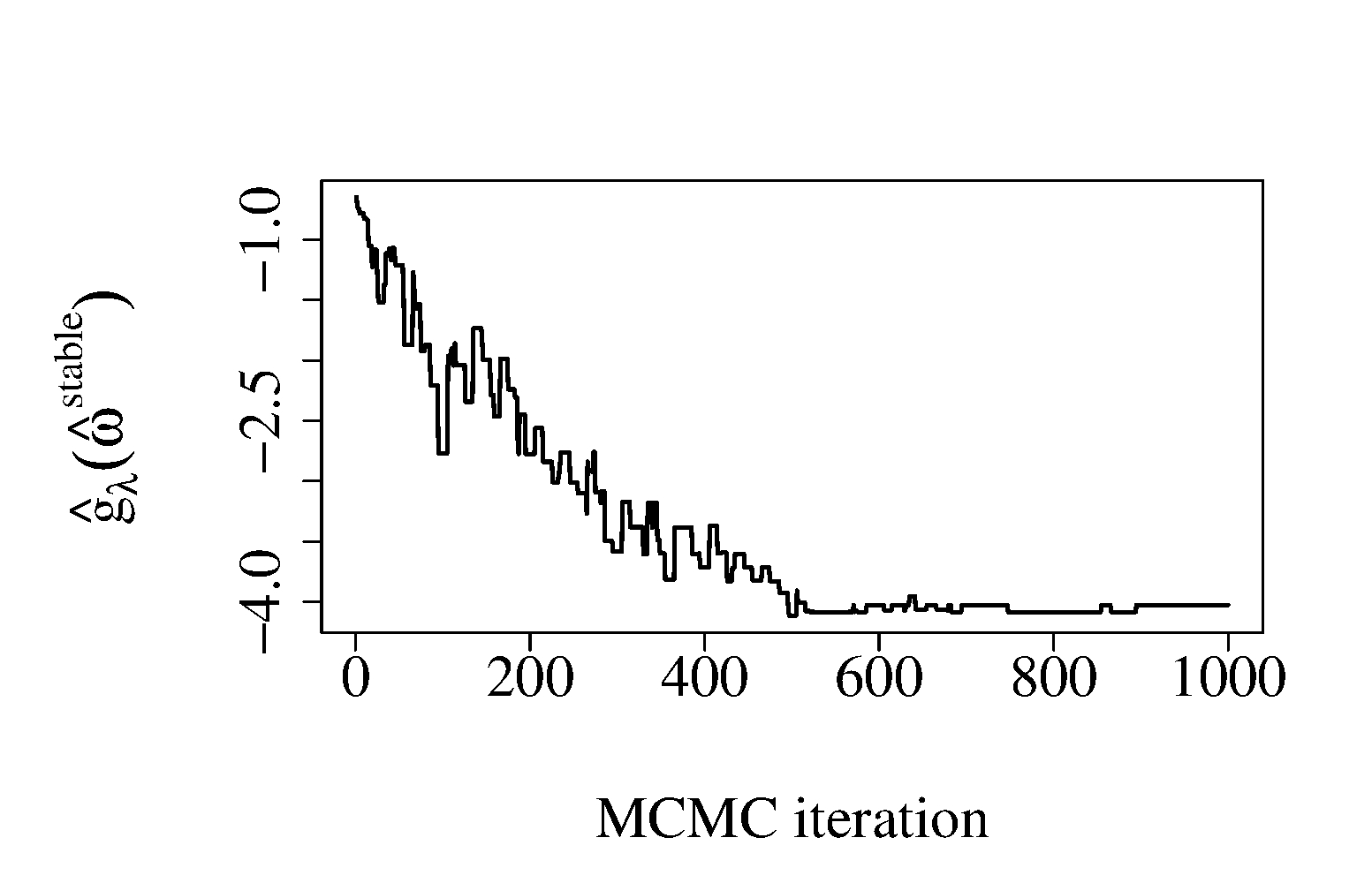}
& \includegraphics[scale=0.4]{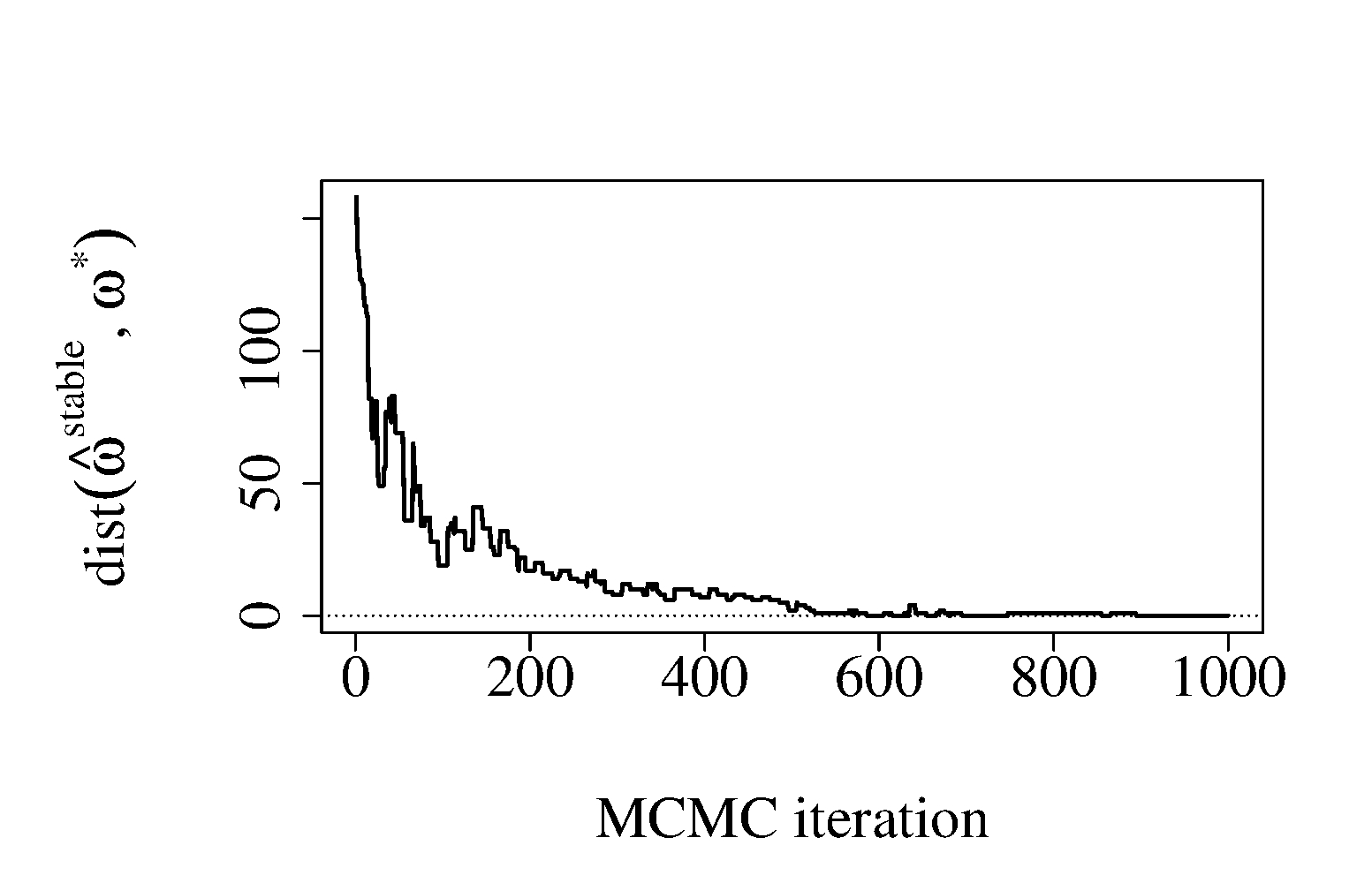}
\end{tabular}
\caption{Stochastic selection for Model 1, $N_d(\mu
\bf1,\bfSigma)$, based on Algorithm 2. (a) Objective function $\ghat_\lambda(\bfomega)$
evaluated at samples drawn from $\pi_{\tau,\lambda}(\bfomega)$; the horizontal solid line is the
control chart limit as described in Section \ref{sec3.2}. (b) Observed frequency of
sampled sub-likelihood components. (c) Estimated objective function evaluated at the progressively
selected likelihood, $\bfomegahat^{\text{stable}}$, based on past samples. (d) Hamming distance
between the progressively selected likelihood, $\bfomegahat^{\text{stable}}$, and the globally
optimal solution, $\bfomega^\ast$. Simulation settings: $n=50$, $d=250$, $d^\ast = 200$ $\tau = d$,
$T=1000$ , burn-in length = $250$, $\alpha = 0.1$, $\lambda=1$.}
\end{figure}

\subsection{Exchangeable Normal Variables with Unknown Correlation} \label{sec5.2}

Let $\bfX \sim N_d(\bf0,\bfSigma(\rho))$, where $\bfSigma(\rho) = \{
(1-\rho) \bfI_d + \rho
{\bf1}_d{\bf1}_d^T\}$ and $0 \leq \rho \leq 1$ is the unknown parameter of interest. The
marginal univariate sub-likelihoods do not contain information on $\rho$, so we consider pairwise
sub-likelihoods
\begin{eqnarray}
\ell_{lm}(\rho) =  - \frac{n}{2} \log (1-\rho^2) - \frac{(SS_{ll}
+ SS_{mm})}{2(1-\rho^2)} + \frac{\rho  SS_{lm}}{1-\rho^2}, \  \ \ 1 \leq l<m \leq d,
\end{eqnarray}
where  $SS_{mm} = \sumin (\xsubi_m)^2$ and $SS_{lm}= \sumin \xsubi_l \xsubi_m$. Given
$\bfomega$, we \replace{solve}{estimate $\rho$ by solving} the \replace{cubic}{composite score}
equation $ 0 =\sum_{j<k} \omega_{jk} U_{jk}(\rho)$ by
Newton-Raphson iterations \replace{using}{where each pairwise score}
\begin{equation}
U_{jk}(\rho) = (1+\rho^2) SS_{jk}  - \rho ( SS_{jj} +  SS_{kk}) + n \rho ( 1-\rho^2)
\end{equation}
\add{is a cubic function of $\rho$.}
It is well known that the composite likelihood estimation can lead to poor results for
this model. \citeasnoun{Cox&Reid04} carries out asymptotic variance
calculations, showing that efficiency losses compared to MLE occur whenever
$d>2$, with more pronounced efficiency losses for $\rho$ near $0.5$. Particularly, if  $d=2$
(exactly one pair), $ARE=1$; if $d>2$, ARE=1 if $\rho$ approaches $0$ or $1$. The next simulation
results show that in finite samples composite likelihood selection is advantageous even in such a
challenging situation.

Since closed-form pairwise score expressions are available for this model, we use the
objective function $\ghat(\bfomega)$ based on the one-step jackknife with pseudo-values
computed as in
Equation (\ref{pseudovalues}).  In Table \replace{2}{3}, we present results from $B=250$ Monte Carlo
runs of the MCMC-CLS algorithm applied to the model with $d = 5, 8, 10$ dimensions, which
correspond to $M={{d}\choose{2}} = 10, 28, 45$ sub-likelihoods, respectively.  We
compute Monte Carlo estimates for the finite-sample relative efficiency $RE =
\widehat{Var}(\hat{\rho}_{apw})/\widehat{Var}(\hat{\rho}_{cl}(\hat{\bfomega}^\ast))$, where $\widehat{Var}$ denotes the Monte Carlo estimate of the finite-sample variance
$\hat{\rho}_{cl}(\hat{\bfomega}^\ast)$ is the estimator selected by the
MCMC-CLS algorithm, while $\hat{\rho}_{apw}$ is the all pairwise (APW) estimator obtained by
including all available pairs (\add{thus} $RE>1$ indicates that our stochastic selection outperforms no
selection). We show values of
$\rho$ around $0.5$, since they correspond to the largest asymptotic efficiency losses of
pairwise likelihood compared to MLE (see \citeasnoun{Cox&Reid04}, Figure 1).
In all considered cases, our stochastic selection method \replace{improved}{improves} the efficiency of the estimator
based on all pairwise components; at the same time, our composite likelihoods employ a much smaller
number of components. For example, when $\rho=0.6$ the  efficiency improvements
range from 8\% to 39\%, using only about half of the available components.

\begin{table}[h] \label{table2}
\centering
\begin{footnotesize}
\begin{tabular}{llcccccccc}
          &&& \multicolumn{3}{c}{$n=10$} && \multicolumn{3}{c}{$n=50$}\\
\hline
           &$M={{d}\choose{2}}=$        &&   10       & 28  & 45 &&   10 & 28  & 45 \\
\hline
\\
$\rho=0.4$ & $RE$      && 1.27(0.03) &  1.11(0.01) & 1.07(0.01)  & &  1.15(0.01) &
1.12(0.01) & 1.06(0.01)\\
            &No. comp.                  && 5.83(0.09) & 13.70(0.16) & 21.12(0.22) & &  7.30(0.08)
&13.82(0.16) & 21.45(0.22)\\
\\
$\rho=0.5$ & $RE$      && 1.36(0.02) & 1.14(0.01) & 1.06(0.01) & &  1.19(0.01) &
1.11(0.01) & 1.06(0.01)\\
            &No. comp.                  && 5.51(0.08) & 13.32(0.07) & 20.96(0.22)  & &  7.01(0.07)&
13.64(0.18)&21.23(0.22)\\
\\
$\rho=0.6$  & $RE$     && 1.39(0.02) & 1.17(0.01) & 1.10(0.01) & & 1.38(0.02)  &
1.14(0.01)& 1.08(0.01)\\
&No. comp.                              && 5.80(0.08) & 12.60(0.16) &20.99(0.20)  & & 5.39(0.08)&
13.27(0.15)& 21.45(0.21)\\
\\
\end{tabular}
\end{footnotesize}
\caption{Pairwise likelihood selection for Model 2, $N_d(\bfzero ,\bfSigma(\rho))$ based on
Algorithm 1: Monte Carlo estimates for: (i) the relative efficiency of the parameter estimate under
selection versus no selection (\replace{$RE = Var(\hat{\rho}_{selection})/Var(\hat{\rho}_{no\_selection})$,}{
$RE =Var(\hat{\rho}_{apw})/Var(\hat{\rho}_{cl}(\hat{\bfomega}^\ast))$,}
so that $RE>1$ indicates that the selection outperforms no selection); (ii) and number of
sub-likelihood components (No. comp.). Monte Carlo standard errors are in parenthesis.
Simulation settings: $\tau =  d$, chain length $T=10d$, $\xi=0.7$.}
\end{table}

\subsection{Real Data Analysis: Australian Breast Cancer Family Study}
\label{example3}

In this section, we apply the MCMC-CLS  algorithm to a real genetic dataset of women
with breast cancer obtained from the Australian Breast Cancer Family Study (ABCFS)
\cite{Dite&al03} and control subjects from the Australian Mammographic Density Twins and Sisters
Study \cite{Odefrey10}. All women were genotyped using a Human610-Quad beadchip array.  The final
data set \add{that we used} consisted of a subset of 20 single nucleotide polymorphisms (SNPs) corresponding
to genes encoding a candidate susceptibility pathway, which is motivated by biological
considerations. After recommended data cleaning and
quality control procedures (e.g., checks for SNP missingness, duplicate relatedness, population
outliers \cite{Weale10}), the final data  consisted of $n=333$ observations ($67$ cases
and $266$ controls).

To detect group effects due to cancer, we consider an extension of the latent multivariate Gaussian
model first introduced by \citeasnoun{Han&Pan12}. Let $\bfY^{(i)} = (Y_{1}^{(i)}, \dots,
Y_{d}^{(i)})$, $i= 1,\dots, n$, be independent observations of a multivariate categorical variable
measured on $n$ subjects. Each variable $Y^{(i)}_{k}$ can take values $0,1$ or $2$, representing the
copy number of one of the
alleles of SNP $k$ of subject $i$. The binary variable $X^{(i)} = x^{(i)} = 0$ or $1$  represents
disease status of the $i$th subject (0 = control and 1 = disease). We assume a latent random
$d$-vector $\bfZ^{(i)} = (Z^{(i)}_{1}, \dots, Z^{(i)}_{d}) \sim N_d(\bfmu^{(i)}(\theta) ,
\bfR)$, where  $\bfmu^{(i)}(\theta)$ is a conditional mean vector with elements
 $\mu_1^{(i)}(\theta)=\cdots= \mu_d^{(i)}(\theta) = \theta x^{(i)}$ and $\bfR$ is the correlation
matrix.  Our main interest is on the unknown mean
parameter $\theta$, which is common to all the SNP variables and represents the main effect due
to disease. We assume $P(Y_{k}^{(i)}=0 | X^{(i)} = x^{(i)}) = P(Z^{(i)}_{k} \leq \gamma_{k1})
\label{latent_model1}$, $P(Y_{k}^{(i)}=1 | X^{(i)} = x^{(i)}) = P(\gamma_{k1}  <  Z^{(i)}_{k} \leq
\gamma_{k2})$, and $
P(Y_{k}^{(i)}=2 | X^{(i)} = x^{(i)}) = P(Z^{(i)}_{k} > \gamma_{k2})$,
where  $\gamma_{k1}$ and $\gamma_{k2}$ are  SNP-specific thresholds. The above model
reflects the ordinal nature of
genotypes and assumes absence of the Hardy-Weinberg equilibrium
(HWE) (allele frequencies and genotypes in a population are constant from generation to generation).
If the HWE holds the parameters $\gamma_{1k}$ and $\gamma_{2k}$ are not needed, since we have the
additional constraint \replace{$P(X_k^{(i)} = 2) = P(X_k^{(i)} = 1)^2$.}{$P(Y_k^{(i)} = 2) = P(Y_k^{(i)} = 1)^2$.}

Let $\bfgamma = \{ (\gamma_{1k}, \gamma_{2k}): k=1,\dots,d \}$ and define intervals
\replace{$\Gamma(Y^{(i)}_{k})$}{$\Gamma_k(Y^{(i)}_{k})$} to \replace{denote intervals}{be}
$(-\infty,\gamma_{k1}]$, $(\gamma_{k1}, \gamma_{k2}]$ and
$[\gamma_{k2}, \infty)$, corresponding to $Y^{(i)}_{k} =0,1$ and $2$, respectively. The full
log-likelihood
function is
\begin{align*}
\ell(\theta, \bfgamma, \bfR) &= \sumin \log
P(\bfY^{(i)}=\bfy^{(i)}| X^{(i)}= x^{(i)}) \\
&= \sumin \log
\int_{\Gamma_1(y^{(i)}_{1})} \cdots \int_{\Gamma_d(y^{(i)}_{d})} f(z_1, \dots, z_d| \bfmu^{(i)}(\theta),
\bfR)
dz_1
\cdots dz_d,
\end{align*}
where $f(z_1, \dots, z_d| \bfmu, \bfR)$ is the pdf of the $d$-variate normal
density function with mean $\bfmu$ and correlation matrix $\bfR$.
Clearly, the full log-likelihood
function is intractable when $d$ is moderate or large, due to the
multivariate integral in the likelihood expression. Note that for the marginal latent
components, we have \replace{$Z_k \sim N_1(0,1)$,}{$Z_k^{(i)} \sim N_1(\theta x^{(i)},1)$,}
so $\bfgamma$ and $\theta$ \replace{can}{can be}
estimated by minimizing the one-wise composite log-likelihood
\begin{align} \label{likSNP2}
\ell_{cl}(\theta, \bfgamma, \bfomega) & = \sum_{k=1}^d \omega_{k} \sumin \log
P(Y_{k}^{(i)}= y_{k}^{(i)}| X^{(i)}= x^{(i)}) \\
& = \sum_{k=1}^d \omega_{k} \sumin
\log \int_{\Gamma_k(y^{(i)}_{k})} \phi(z_k|\theta x^{(i)},1) dz_k,
\end{align}
where $\phi(\cdot|\mu,1)$ denotes the normal pdf with mean $\mu$ and unit variance.
\add{We focus on using the one-wise composite log-likelihood in this section, except
when $\bfR$ is to be estimated where we use pairwise composite log-likelihood.}
Differently from the expression in \citeasnoun{Han&Pan12}, the
disease group effect $\theta$ is common to multiple sub-likelihood components; also, we allow
for the inclusion/exclusion of particular sub-likelihood components (corresponding to SNPs) by
selecting
$\bfomega$.
\begin{figure}[h] \label{figure3}
\centering
\begin{tabular}{cc}
(a) & (b) \\
 \includegraphics[scale=0.4]{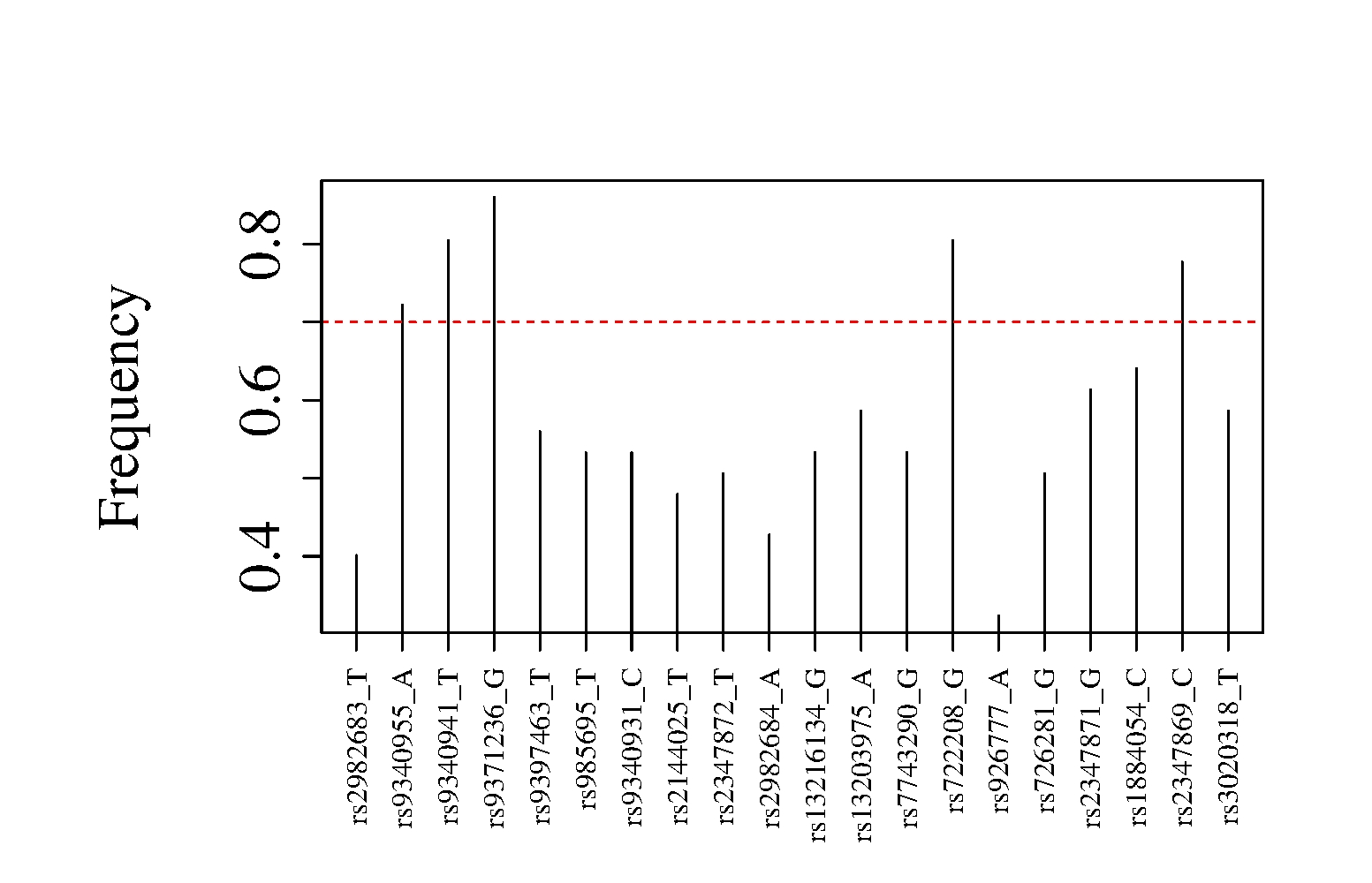}
&\includegraphics[scale=0.4]{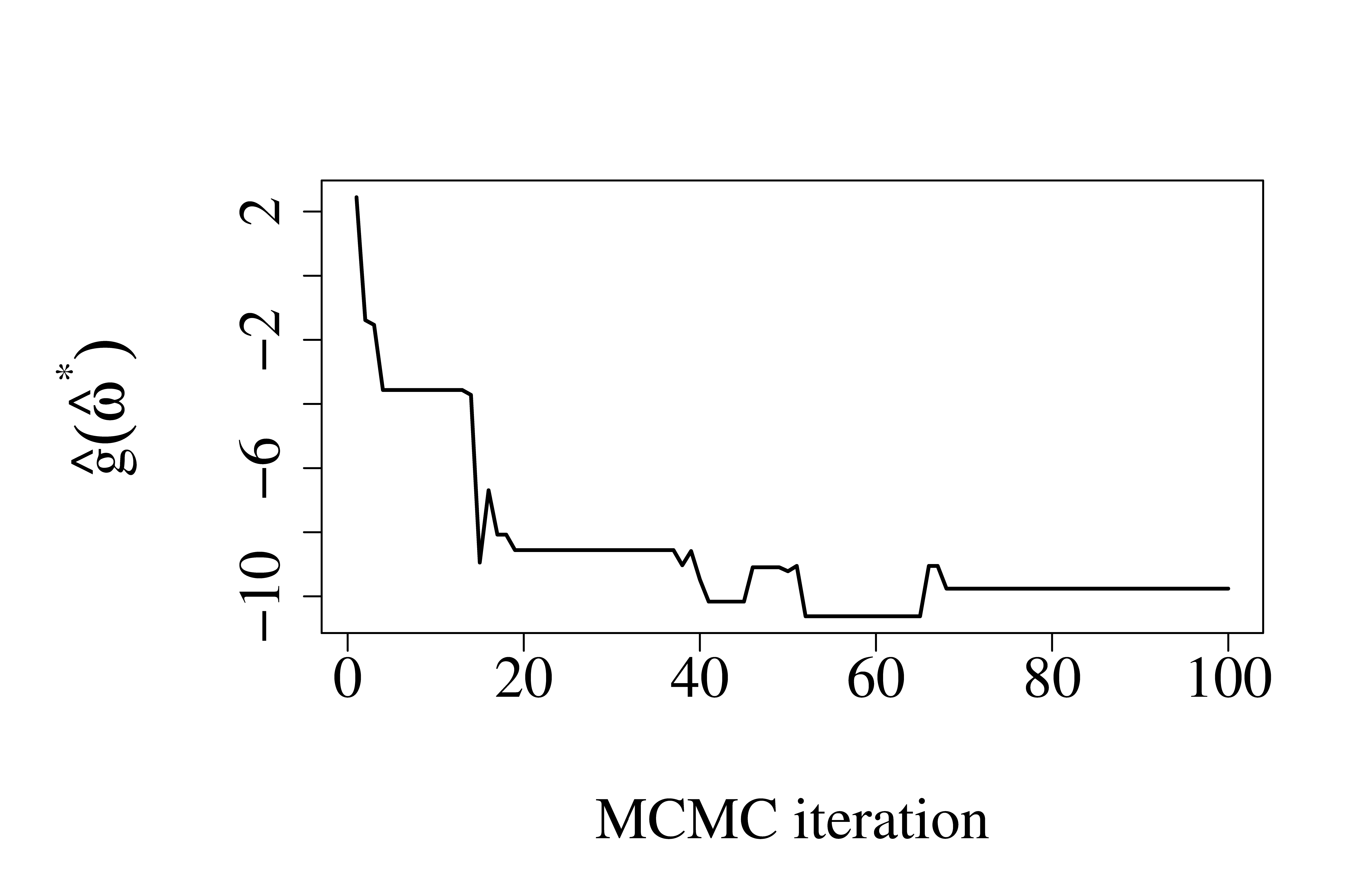}\\
(c) & (d) \\
\includegraphics[scale=0.4]{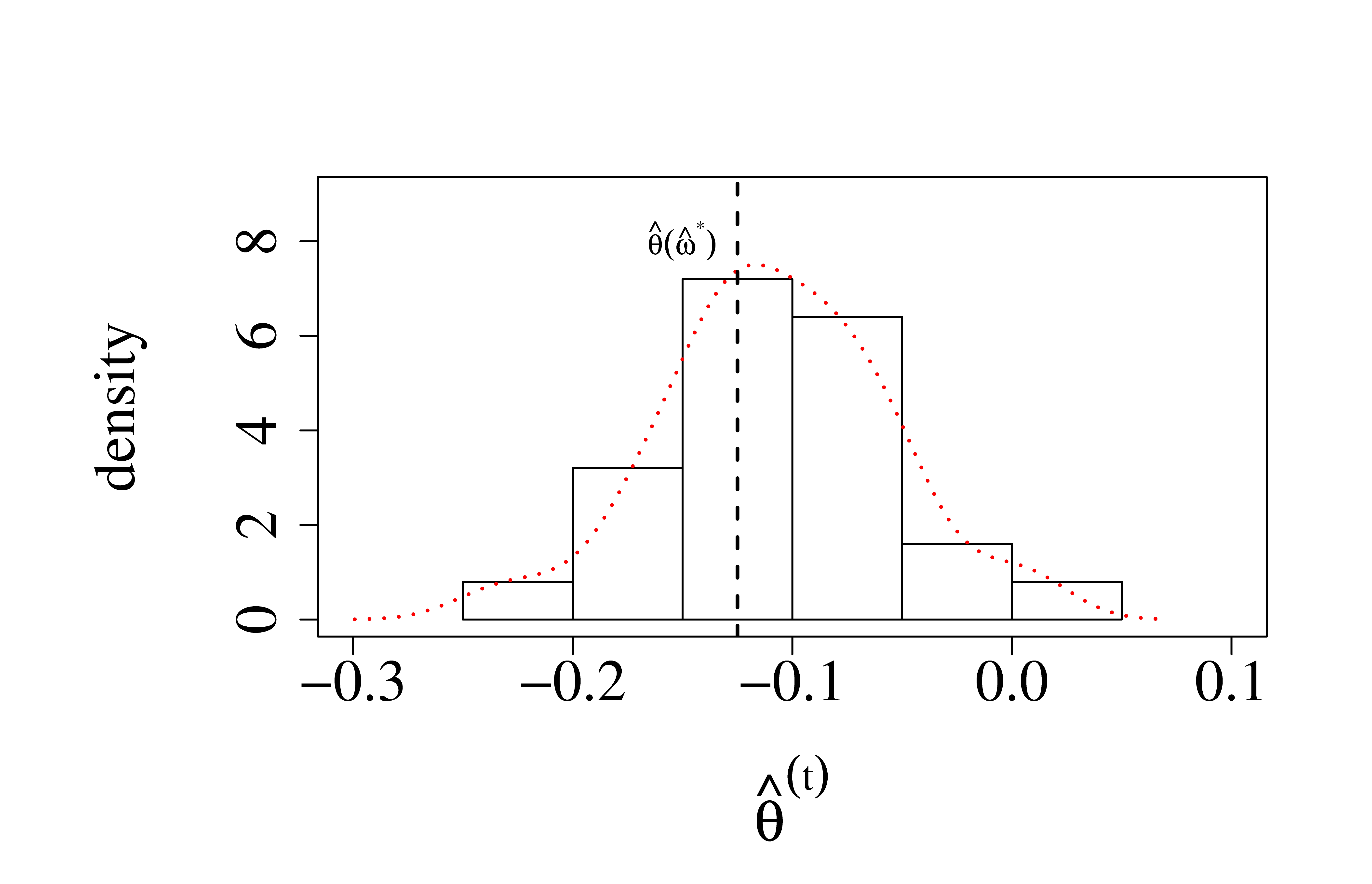}
& \includegraphics[scale=0.4]{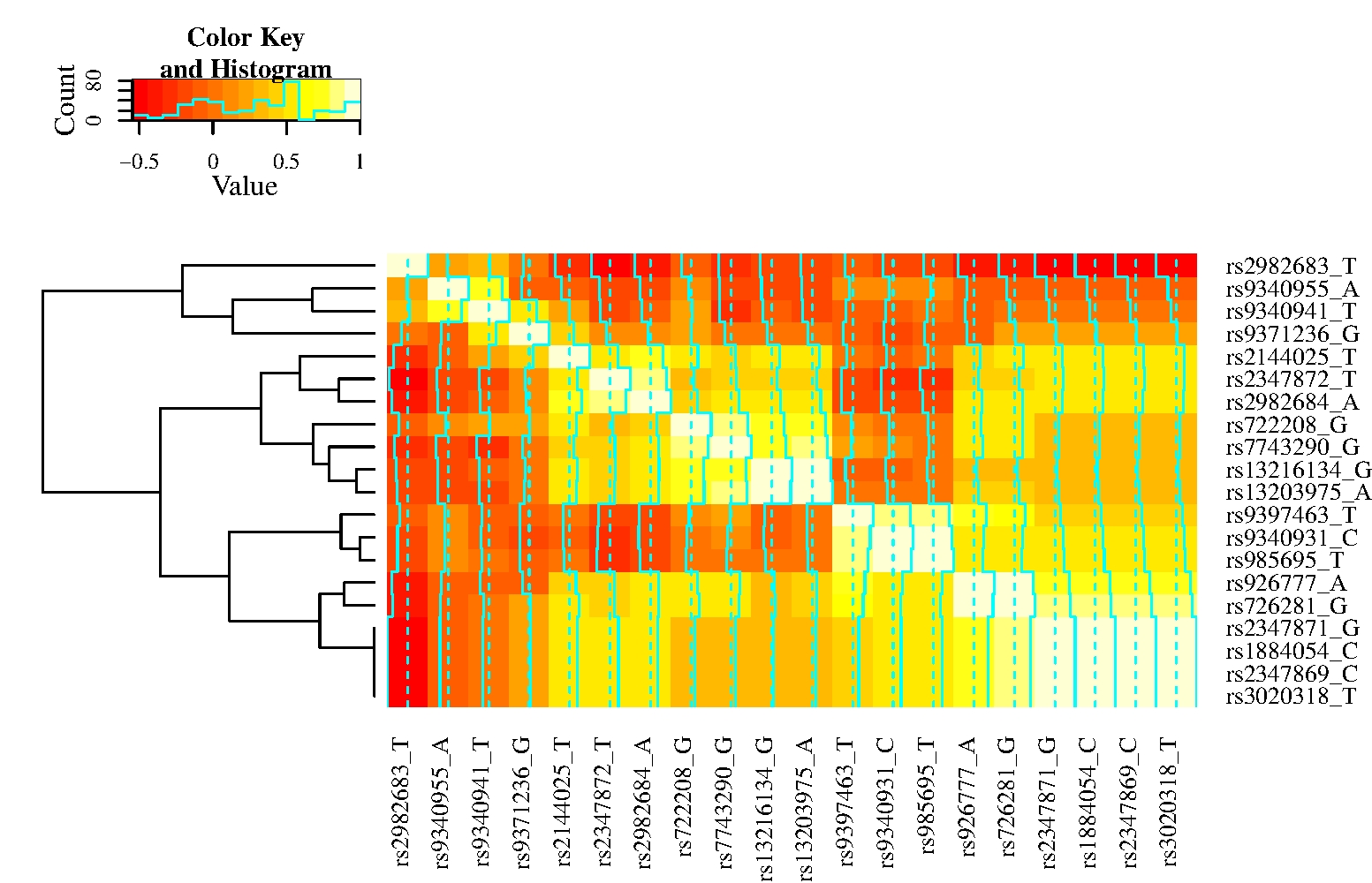}
\end{tabular}
\caption{Composite likelihood selection for the ABCFS data by Algorithm 1: (a) Frequency of
the sampled marginal likelihood components; (b) Objective function evaluated at the current
solution $\bfomegahat^\ast$ (computed from past samples with $\xi=0.7$); (c) Parameter estimates,
$\hat{\theta}^{(t)} = \hat{\theta}(\bfomega^{(t)})$, based on sampled composition rules,
$\bfomega^{(t)}$ with optimal parameter estimate $\hat{\theta}(\bfomega^\ast)$ corresponding to
vertical dashed line. (d) Pairwise likelihood estimates for the correlation matrix $\bfR$ for SNPs
in the susceptibility pathway.}
\label{fig2}
\end{figure}
We estimated the optimal composition rule $\hat{\bfomega}^\ast$ based on Gibbs samples from
Algorithm 1, where the objective function $\ghat(\bfomega)= \log Var(\hat{\theta}(\bfomega))$ was
estimated by delete-10  jackknife. We selected five marginal likelihoods (SNPs) occurring
with at least $\xi=0.7$ frequency in 250 runs of the Gibbs sampler (see Figure 2 (a)).
In Figure 2 (b), we show the trajectory of the objective function $\ghat(\bfomega)$ evaluated at the
 current optimal solution  $\bfomegahat^\ast$ (optimal solutions are computed from past
samples using a  $\xi=0.7$ threshold).  The estimated variance tends to \replace{decrease}{sway toward
the minimum} as more composition
rules are drawn by our Gibbs sampler. This behavior is in agreement with preliminary simulation
results carried out on this model (not presented here) as well as the \replace{example}{examples} presented in
Sections \ref{sec5.1} and \ref{sec5.2}.

Figure 2 (c) shows the empirical distribution of parameter estimates,
$\hat{\theta}^{(t)} = \hat{\theta}(\bfomega^{(t)})$, based on sampled
vectors $\bfomega^{(t)}$, $t=1,\dots, 250$. The vertical dashed line  corresponds to the
selected parameter estimate
$\hat{\theta}(\hat{\bfomega}^\ast)$, which is located near the mode of the
empirical distribution.  Particularly, the selected McLE is
$\hat{\theta}(\hat{\bfomega}^\ast) = -0.125$ and the corresponding delete-10 jackknife standard
error is $\hat{sd}(\hat{\theta}(\hat{\bfomega}^\ast))=0.012$. The McLE based on \replace{all}{using all} 20
target SNPs is $\hat{\theta}_{all} = -0.112$ with the \remove{correspondent} delete-10 jackknife standard error
$\hat{sd}(\hat{\theta}_{all})=  0.042$. Our estimator \replace{gave}{gives} a
substantial accuracy improvement, supporting the conclusion of a difference between case and control
groups (i.e., $\theta \neq 0$) with higher confidence. Finally, in Figure 2 (d), we show
estimates for the correlation matrix $\bfR$ for the target SNPs, based on
the pairwise composite likelihood described in \citeasnoun{Han&Pan12}.

\section{Final remarks} \label{sec:remarks}

Composite likelihood estimation is a rapidly-growing need for a number of fields, due
to the astonishing growth of data complexity and the limitations of traditional maximum
likelihood
estimation in this context.  The Gibbs sampling protocol proposed in this paper
addresses an important unresolved issue by providing a tool to automatically select the
most useful
sub-likelihoods from a pool of feasible components. Our numerical results on
simulated and real data show that the composition rules generated by our  MCMC approach are useful
to improve the variance of traditional McLE estimators, \replace{which typically include all the
available components.}{typically obtained by using all sub-likelihood components available.}
Another advantage deriving from our method is the possibility to  generate
sparse composition rules, since our Gibbs sampler selects only a (relatively  small)
subset of informative sub-likelihoods while discarding non-informative or redundant
components.

In the present paper, likelihood sparsity derives naturally from the discrete
nature of our MCMC approach based on  binary composition rules with \replace{$\bfomega \in \{0,1
\}^d$.}{$\bfomega \in \{0,1\}^M$.} On the other hand, the development  of sparsity-enforcing
likelihood selection methods suited to
real-valued weights would be valuable as well and could result in more efficient composition
rules. For example, \citeasnoun{Lindsay&al2011} discuss the optimality of composite likelihood
estimating functions involving both positive and negative real-valued weights.
Actually, the row averages $\overline{\omega}_1, \cdots, \overline{\omega}_M$ computed in
Step~2 of Algorithm~1 can also be used to replace $\bfomega=(\omega_1,\cdots,\omega_M)^T$
in \replace{$\ellhat_{cl}(\bftheta, \bfomega)$}{$\ell_{cl}(\bftheta, \bfomega)$} defined by (\ref{comp_lik}), providing a composite
log-likelihood with between-0-and-1 weights. It would be of interest to see how well this
form of composite log-likelihood gets on in regard to the efficiency. This, however, was
not pursued in this paper and is left for future exploration.
Finally, the penalized version of the objective function described in Section
\ref{sec4} enforces \remove{arbitrarily} sparse likelihood functions, which is necessary in situations where
the model
complexity is relatively large compared to  the sample size. Thus developing a thorough theoretical
understanding on the effect of the penalty on the selection as $d,n \rightarrow \infty$ would be
very valuable for improved selection algorithm in high-dimensions.

\appendix

\makeatletter   
 \renewcommand{\@seccntformat}[1]{APPENDIX~{\csname the#1\endcsname}.\hspace*{1em}}
 \makeatother

\bibliographystyle{ECA_jasa}
\bibliography{biblio}

\end{document}